\documentclass[sigconf,screen]{acmart}
\usepackage{booktabs}
\usepackage{multirow}
\usepackage{enumitem}
\usepackage{tcolorbox}
\usepackage{xspace}

\usepackage[]{collab}
\collabAuthor{zb}{blue}{Zhuangbin}
\collabAuthor{zx}{teal}{Zhixiang}

\usepackage{seqsplit} %长字符串换行且不改变字体 \seqsplit

\usepackage{soul} % 防止下划线里面换行 \ul

\newcommand{\BugName}{iBug\xspace}
\newcommand{\BugNames}{iBugs\xspace}
\newcommand{\BugNumber}{121\xspace}
\newcommand{\FullBugNumber}{364\xspace}
\newcommand{\ImpactNumber}{86 \xspace}

\newcommand{\SoftwareInteraction}{Intra-system Interaction}
\newcommand{\Softwareinteraction}{Intra-system interaction}
\newcommand{\softwareinteraction}{intra-system interaction}
\newcommand{\HardwareInteraction}{Hardware Interaction}
\newcommand{\Hardwareinteraction}{Hardware interaction}
\newcommand{\hardwareinteraction}{hardware interaction}
\newcommand{\EnvironmentInteraction}{Environmental Interaction}
\newcommand{\Environmentinteraction}{Environmental interaction}
\newcommand{\environmentinteraction}{environmental interaction}

\newcounter{finding}

\setcopyright{acmlicensed}
\copyrightyear{2025}
\acmYear{2025}
\acmDOI{XXXXXXX.XXXXXXX}

\acmISBN{978-1-4503-XXXX-X/18/06}

\begin{document}

% \title{Understanding Interaction Bugs in Robotic Systems}
\title{An Empirical Study of Interaction Bugs in ROS-based Software}

\author{Zhixiang Chen}
\affiliation{
  \institution{Sun Yat-Sen University}
  \department{School of Software Engineering}
  \city{Zhuhai}
  \country{China}}
\email{chenzhx69@mail2.sysu.edu.cn}

\author{Zhuangbin Chen*}
\affiliation{
  \institution{Sun Yat-Sen University}
  \department{School of Software Engineering}
  \city{Zhuhai}
  \country{China}}
\email{chenzhb36@mail.sysu.edu.cn}

\author{Xingjie Cai}
\affiliation{
  \institution{Sun Yat-Sen University}
  \department{School of Software Engineering}
  \city{Zhuhai}
  \country{China}}
\email{caixj27@mail2.sysu.edu.cn}

\author{Wei Li}
\affiliation{
  \institution{Sun Yat-Sen University}
  \department{School of Software Engineering}
  \city{Zhuhai}
  \country{China}}
\email{liwei378@mail2.sysu.edu.cn}

\author{Zibin Zheng}
\affiliation{
  \institution{Sun Yat-Sen University}
  \department{School of Software Engineering}
  \city{Zhuhai}
  \country{China}}
\email{zhzibin@mail.sysu.edu.cn}

% \affiliation{%

% \affiliation{%



\begin{abstract}

Modern robotic systems integrate multiple independent software and hardware components, each responsible for distinct functionalities such as perception, decision-making, and execution. These components interact extensively to accomplish complex end-to-end tasks. As a result, the overall system reliability depends not only on the correctness of individual components, but also on the correctness of their interactions. Failures often manifest at the boundaries between components, yet interaction-related reliability issues in robotics—referred to here as interaction bugs (iBugs)—remain underexplored.

This work presents an empirical study of iBugs within robotic systems built using the Robot Operating System (ROS), a widely adopted open-source robotics framework. A total of 121 iBugs were analyzed across ten actively maintained and representative ROS projects. The identified iBugs are categorized into three major types: intra-system iBugs, hardware iBugs, and environmental iBugs, covering a broad range of interaction scenarios in robotics.
The analysis includes an examination of root causes, fixing strategies, and the impact of these bugs. Several findings are derived that shed light on the nature of iBugs and suggest directions for improving their prevention and detection. These insights aim to inform the design of more robust and safer robotic systems.
\end{abstract}

%    <concept>
%        <concept_id>10002944.10011123.10010912</concept_id>
%        <concept_desc>General and reference~Empirical studies</concept_desc>
%        <concept_significance>500</concept_significance>
%        </concept>
%    <concept>
%        <concept_id>10010520.10010553.10010554</concept_id>
%        <concept_desc>Computer systems organization~Robotics</concept_desc>
%        <concept_significance>500</concept_significance>
%        </concept>
%    <concept>
%        <concept_id>10002944.10011123.10010577</concept_id>
%        <concept_desc>General and reference~Reliability</concept_desc>
%        <concept_significance>500</concept_significance>
%        </concept>
%  </ccs2012>

% \ccsdesc[500]{General and reference~Empirical studies}
% \ccsdesc[500]{Computer systems organization~Robotics}
% \ccsdesc[500]{General and reference~Reliability}
% \keywords{Empirical study, Reliability, Robotic systems, Interaction bugs}

\setcopyright{none} % to remove the copyright notice
\settopmatter{printacmref=false} % to remove the ACM Reference Format
\renewcommand\footnotetextcopyrightpermission[1]{} %remove the entire box in the bottom left


\maketitle

\section{Introduction}  
\label{sec:intro}

% 介绍机器人的应用广泛，市场有多少钱
% Robots are increasingly being implemented across a wide range of industries due to their versatility and efficiency. From manufacturing and healthcare to agriculture and logistics, robots improve productivity and precision. This widespread adoption has led to significant growth in the robotics market, with the market size expected to reach \$46.11 billion in 2024 and continue growing at an annual rate of 9.63\% \cite{RoboticMarket}.

Robots are becoming an integral part of our society, revolutionizing industries and enhancing everyday life.
In the realms of manufacturing, healthcare, and logistics, robots can handle repetitive, hazardous, or complex tasks, significantly improving productivity and safety~\cite{faheem2024ai-advantage,anumandla2023robotic-advantage,javaid2021substantial-advantage}.
% From manufacturing and healthcare to agriculture and logistics, robots improve productivity and precision.
This widespread adoption has led to substantial growth in the robotics market.
The market size has reached
% \$71,196.4 million
\$71.2 billion in 2023 and is projected to grow to 
% \$286,798.3 million
\$286.8 billion by 2032, with a compound annual growth rate of 18.4\% \cite{RoboticMarket}.

% It is anticipated that the market size will reach \$46.11 billion in \red{2024}\zb{we already in 2024}, with an annual growth rate of 9.63\%~\cite{RoboticMarket}.

% 介绍机器人系统，需要实现多少的功能，具有复杂、庞大的体系，里面需要处理大量交互，里面存在三种交互的正确处理是有挑战性的，因此容易出现各种bug，举三个例子

% Modern robotic systems consist of a series of sophisticated modules, each responsible for specific functionality like sensor data processing, path calculating, motor actuation, etc. These modules work collaboratively to enable robots to achieve automation and intelligent operations in various application scenarios. Each module in the robotic system can be easily reused even if it is usually developed and maintained by different developers with extensive domain knowledge.

Modern open-source frameworks for robotic system development, such as ROS~\cite{quigley2009ros}, Player~\cite{DBLP:conf/iros/GerkeyVSHSM01-player}, Orca~\cite{DBLP:conf/icra/BrooksKMWO05-Orca}, and YARP~\cite{metta2006yarp}, often employ a modular architecture~\cite{luckcuck2022compositional-modular-robotic,DBLP:journals/computers/JahnWS19-modular-robotic}, enabling the entire system to be built from interoperable components~\cite{DBLP:conf/iros/BrooksKMWO05-component-based-robotics,DBLP:journals/ram/BrugaliS09-component-based-robotics,DBLP:journals/ram/BrugaliS10-component-based-robotics}.
% which allows the entire system to be built from interoperable components~\cite{DBLP:conf/iros/BrooksKMWO05-component-based-robotics,DBLP:journals/ram/BrugaliS09-component-based-robotics,DBLP:journals/ram/BrugaliS10-component-based-robotics}.
% which promotes code reusability and simplifies the creation of advanced robotic software.
% Modern open-source frameworks for developing robotic systems, e.g., ROS \cite{quigley2009ros}, Player \cite{DBLP:conf/iros/GerkeyVSHSM01-player}, Orca \cite{DBLP:conf/icra/BrooksKMWO05-Orca}, and YARP \cite{metta2006yarp}, employ a modular design where the entire system is composed of interoperable components to facilitate code reuse and ease the development of sophisticated robotic software.
Known as the ``Linux of Robotics,'' the Robot Operating System (ROS) stands as the \textit{de facto} choice in this field~\cite{DBLP:conf/icsm/KolakAGHT20-ecosystem,DBLP:conf/icse/DurschmidTGG24-ROSInfer,DBLP:conf/icse/MalavoltaLSLG20-architect}.
Its package ecosystem provides developers with reusable, off-the-shelf components that encapsulate common robot functions (e.g., perception, navigation)~\cite{DBLP:journals/jss/EstefoSRF19-package-reuse,DBLP:conf/sigsoft/00020BBP20-service-robot,DBLP:conf/etfa/AwadHRB16-workbench}.
The hardware-agnostic nature of ROS also allows for the flexible assembly of various hardware elements, e.g., sensors and actuators.
Such modular design empowers developers to rapidly prototype robotic systems by integrating software and hardware components, with appropriate configurations to match their intended application and environment.
In ROS-based systems, different components are engaged in comprehensive interactions to execute seamless, end-to-end operations efficiently.
For instance, the navigation unit formulates a path based on environmental data gathered by sensors, and subsequently instructs the actuators to perform physical motions.
Such intricate interactions fundamentally differ from traditional, static distributed systems, such as web-based client-server architectures or high-performance computing clusters.
Robotic systems demand real-time coordination of software, hardware, and environmental dynamics, as well as continuous sensor-feedback loops within unpredictable physical conditions.
% \zb{try to remove one line}
% real-time constraints and continuous sensor-feedback loops
% by their tight integration of software, hardware, and environmental dynamics. They require real-time coordination under unpredictable physical conditions, while managing safety-critical specifications and heterogeneous hardware components---challenges rarely faced in traditional software-only modular systems.
%  safety-critical specifications and environmental perturbations  
% \zx{Here say ``modular'' system or ``distributed'' system? ``distributed'' is not mentioned before}
% \zb{the interaction in robotic systems is more complex and dynamic than traditional distributed systems (which is fixed, robots can move), so studying iBugs is important}

Despite the benefits of modularization, bugs can reside in the interactions of multiple separate components, which we term as \textit{interaction bugs} (or \textit{\BugNames}) in this study.
Such \BugNames pose significant threats to the reliability of robotic systems, and even their safety.
For example, a cancel request from the user interface unexpectedly triggered a crash in the control module, causing the robot to continue moving even after colliding with an obstacle \cite{NavIssue-continues-moving-when-cancel}.
This can be attributed to the increasing complexity of robots' capabilities and the heterogeneity of hardware.
The fact that every (sub-)system tends to be developed and maintained independently by different projects and teams further compounds the problem~\cite{DBLP:conf/icsm/KolakAGHT20-village}.
Therefore, building correct and dependable robotic systems becomes a challenging task.
Developers must tackle not only the bugs within individual components, but also the potential \BugNames that could arise from the numerous combinations of different components.

% Substantial \zb{empirical} empirical studies \cite{DBLP:conf/icse/Fischer-Nielsen20-ROSDependency,DBLP:conf/issta/CanelasSFT24-misconfiguration,DBLP:journals/ese/TimperleyHSDW24-ROBUST221bugs} have investigated the reliability of robotic systems in the literature. 
Some empirical studies \cite{DBLP:conf/icse/Fischer-Nielsen20-ROSDependency,DBLP:conf/issta/CanelasSFT24-misconfiguration,DBLP:journals/ese/TimperleyHSDW24-ROBUST221bugs} have been conducted to investigate the reliability of robotic systems in the literature. 
For example,~\cite{DBLP:conf/icse/Fischer-Nielsen20-ROSDependency} examines bugs arising from incorrect references to external dependencies, while~\cite{DBLP:conf/issta/CanelasSFT24-misconfiguration} focuses on misconfigurations primarily related to configuration files (e.g., launch files, YAML-based parameter settings). 
Furthermore, ~\cite{DBLP:journals/ese/TimperleyHSDW24-ROBUST221bugs} investigates general bugs in ROS-based systems, categorizing bug faults into seven coarse-grained types.
% \zx{need this sentence?} 
% For example, one study has examined software faults arising from incorrect references to external software assets, termed as dependency bugs \cite{DBLP:conf/icse/Fischer-Nielsen20-ROSDependency}. Another study has investigate misconfigurations that primarily relates to configuration files, such as launch XML and parameter YAML files \cite{DBLP:conf/issta/CanelasSFT24-misconfiguration}. Besides, general bugs across seven ROS-based systems have been studied 
% For instance, researchers have conducted empirical studies on general types of bugs \cite{DBLP:conf/icse/Fischer-Nielsen20-ROSDependency,DBLP:journals/ese/TimperleyHSDW24-ROBUST221bugs,DBLP:conf/issta/CanelasSFT24-misconfiguration} and vulnerabilities \cite{DBLP:conf/compsac/CottrellBSR21-ROSvul} in robotic systems.
% Static \cite{DBLP:conf/sigsoft/KateCCZE21-physframe,DBLP:conf/icsa/TimperleyDSGG22-ROSDiscover,DBLP:conf/icse/DurschmidTGG24-ROSInfer} and dynamic \cite{ROS2Fuzz,DBLP:conf/icra/XieBZ022-ROZZ-Fuzz,DBLP:conf/sigsoft/KimK22-RoboFuzz,DBLP:conf/asplos/BaiS024-ROFER-Fuzz,DBLP:conf/issta/ShenLXSWGS024-FuzzingCallback} bug detection tools have also been proposed.
% 现有工作都主要关注软件内部的Bug，
% However, none of these studies has focused on \BugNames, a critical failure mode in component-based systems~\cite{DBLP:conf/eurosys/TangBZKJGX23}.
However, \BugNames, a critical failure mode in component-based systems \cite{DBLP:conf/eurosys/TangBZKJGX23}, has remained largely underexplored in existing studies. 
% \blue{Given that robotic systems operate in physical environments, \BugNames can lead to severe consequences, including system failures, unpredictable behavior, or even physical harm.}
Given that robotic systems operate in physical environments, \BugNames can lead to severe consequences (e.g., physical harm), making its understanding paramount to researchers and practitioners. This understanding can facilitate the identification of areas (e.g., detection and diagnosis) requiring better tool support and provide valuable insights for the development of high-quality robotic systems.
% The understanding of \BugNames is of paramount significance to both researchers and practitioners in the robot community, which can facilitate the identification of areas (e.g., detection and diagnosis) requiring better tool support and provide valuable insights for the development of high-quality robotic systems. 
% \zb{elaborate the differences with existing work}
% It provides valuable insights for the design of development and diagnostic tools to facilitate the building of high-quality robotic systems.
% 目前针对机器人软件的研究有xxxx，都干了xxx，但是还没有针对交互bug的研究。研究交互Bug对机器人软件的可靠性具有重要意义

In this work, we conduct the \textit{first} empirical study on the \BugNames of robotic systems based on ROS.
% We select nine representative projects from 652\zb{same problem here, 9 over 652} active repositories in the official ROS index for our study.
We select ten representative projects from the active repositories in the official ROS index \cite{ROS-Index} for our study.
We identify \FullBugNumber \BugNames through keyword searches and manual confirmation, from which we randomly sample 121 \BugNames for further investigation.
% \red{We investigate 1,160 issues (including pull requests) with a bug label and a closed status from the selected repositories and finally obtain 75 \BugNames.}
Through a detailed analysis of these \BugNames, we try to answer the following three research questions.

\begin{itemize}[noitemsep,leftmargin=5.5mm]
    \item \textbf{RQ1 (Root causes)}: What are the root causes of \BugNames?
    \item \textbf{RQ2 (Fixes)}: How do developers fix \BugNames, and are there any common fixing patterns?
    \item \textbf{RQ3 (Bug impacts)}: What is the impact of \BugNames on the behavior of robots?
\end{itemize}

We categorize the collected \BugNames into three types, i.e., \textit{intra-system \BugNames}, \textit{hardware \BugNames}, and \textit{environmental \BugNames}, which reflect the comprehensive range of interaction scenarios during the operations of robotic systems. 
Specifically, intra-system iBugs are defects related to component interactions within the ROS system, hardware iBugs involve malfunctions or miscommunications with physical components, and environmental iBugs arise from unexpected external conditions or internal mishandling in related steps (e.g., collision detection), compromising the robot's correctness or performance.
% or mishandling of ... affecting the robot's correctness or performance.
Through our in-depth analysis of the \BugNames, we have obtained several interesting findings.
For instance, 19.01\% of iBugs are associated with interaction topology problems, which result from developers incorrectly setting up the interaction structure of ROS nodes and interfaces.
Additionally, developers can mismanage hardware resources (e.g., incorrect allocation and release), leading to 50\% of hardware iBugs.
% For example, when multiple messages arrive at a node simultaneously, their callbacks may execute concurrently, leading to concurrency issues where one update to a variable overwrites another.
% Moreover, the intricate physical environment can trigger latent bugs in algorithms, leading to unexpected robot behaviors, such as walking down a wrong path.
% \blue{For example, when multiple messages arrive at a single node simultaneously, the associated callbacks may execute simultaneously and access the same variable. Concurrency issues can arise consequently, where one of the updates to the variable is overwritten by another. }\zb{what's the problem of this case? concurrency?}
For \BugNames under the same root cause, we summarize 22 frequently-used patterns for fixing, covering 76.19\% of the atomic root causes and 60.33\% of the \BugNames, indicating that a substantial portion of \BugNames can be addressed through reusable fixing strategies.
% However, more than half of the bugs are still addressed through case-by-case fixes. This indicates that the resolutions of \BugNames are complex and require a thorough investigation of the specific scenarios of the \BugNames.
Based on our findings, we provide valuable insights for practitioners and researchers in terms of \BugNames avoidance and detection.
For instance, developers should pay more attention to the management of intra-system interactions in terms of topology structure issues, executor misuse, and interface misuse.
Researchers are suggested to develop static methods to detect hardware \BugNames based on the inconsistencies between the implementation and documentation of hardware interfaces.
The studied \BugNames are publicly available on Github \cite{our_dataset}.
% \href{https://anonymous.4open.science/r/Understanding-Interaction-Bugs-in-Robotic-Systems-FC8E/README.md}{https://anonymous.4open.science/r/Understanding-Interaction-Bugs-in-Robotic-Systems-FC8E}.

The main contributions of this work are summarized as follows.

\begin{itemize}[noitemsep,leftmargin=5.5mm]
    \item We conduct the first systematic empirical study on \BugNames within the ecosystem of ROS, which is the most prevalent open-source framework for robotic system development.
    \item We provide a dataset consisting of representative \BugNames in robotic systems, which aims to facilitate a deeper understanding of \BugNames among researchers and practitioners.
    \item We discuss practical directions for \BugName avoidance and detection, which serve as important guidelines for the community to navigate the complexities of robotic system interactions and enhance overall system reliability and safety. 
\end{itemize}

\section{Background}
\label{sec:background}

\subsection{Robot Operating System (ROS)}
\label{subsec:Background-ROS}

ROS is an open-source framework with a set of tools, libraries, and conventions to facilitate the development of complex and reliable robotic systems~\cite{quigley2009ros}.
It provides services designed for a heterogeneous computer cluster such as hardware abstraction, implementation of commonly-used functionality, message-passing between processes, etc.
ROS has attracted numerous developers and become the \textit{de facto} standard for robotic system development~\cite{DBLP:conf/icsm/KolakAGHT20-ecosystem,DBLP:conf/icse/DurschmidTGG24-ROSInfer,DBLP:conf/icse/MalavoltaLSLG20-architect}.

% Robot operating system (ROS) is an open-source framework for developing robotic software~\cite{quigley2009ros}. 
% With its modular and distributed features, ROS allows developers to rapidly develop robotic software by utilizing the rich software packages within its ecosystem without reinventing the wheel~\cite{DBLP:conf/icsm/KolakAGHT20-ecosystem}.
% ROS has attracted a vast number of developers and become the de facto standard for robotic software~\cite{DBLP:conf/icsm/KolakAGHT20-ecosystem,DBLP:conf/icse/DurschmidTGG24-ROSInfer,DBLP:conf/icse/MalavoltaLSLG20-architect}.

\begin{figure*}
    \centering
    \includegraphics[width=0.74\textwidth]{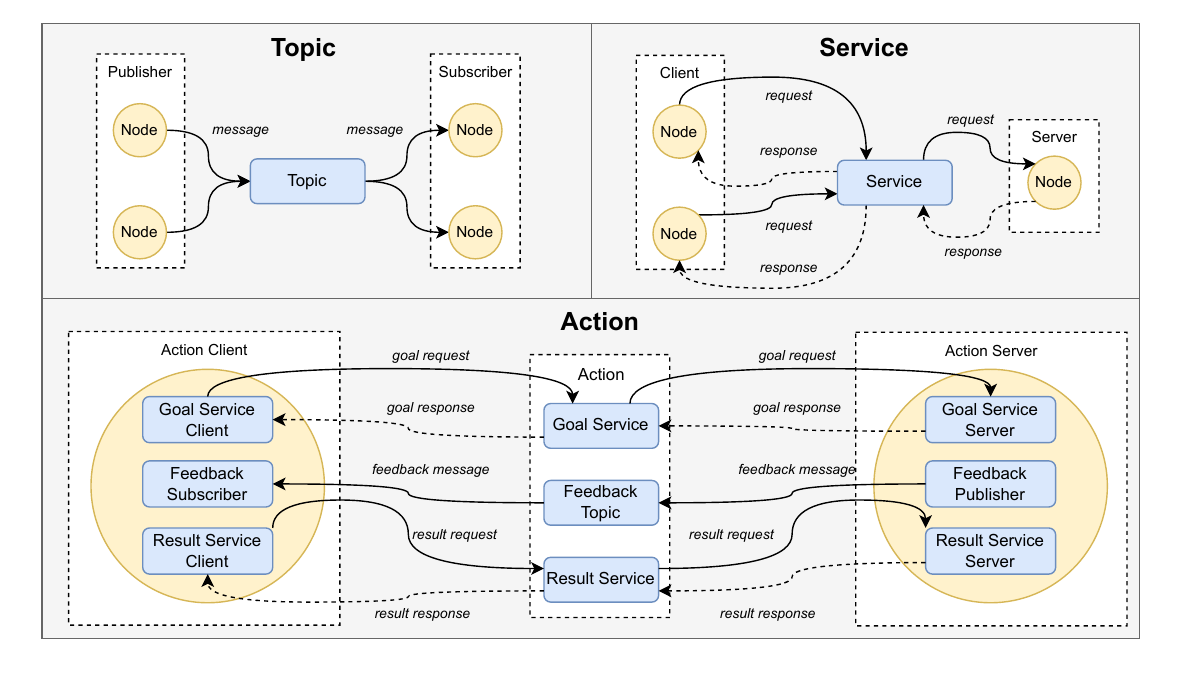}
    \caption{Three interfaces of interaction between nodes in ROS.}
    \label{fig:ROS-node-graph}
\end{figure*}

ROS follows a modular architecture which enables developers to build systems by integrating a diverse array of software packages from its ecosystem~\cite{DBLP:conf/icsm/KolakAGHT20-ecosystem}.
Typical categories of packages include \textit{perception} (e.g., sensor data processing for cameras and LiDARs, environment interpretation), \textit{navigation} (e.g., path planning, localization, and obstacle avoidance), \textit{control} (e.g., joint control, trajectory planning, and inverse kinematics computation), etc.
The fundamental building block of a ROS-based system is \textit{Node}, which is an independent process responsible for executing one of these specific tasks.
For a robot to execute complex operations, its nodes must be able to communicate with each other seamlessly.
As shown in Figure~\ref{fig:ROS-node-graph}, the communication framework of ROS encompasses three primary types of interfaces, i.e., \textit{Topic}, \textit{Service}, and \textit{Action}, each tailored for different interaction scenarios.
The data transmitted through these interfaces is pre-defined and must conform to the corresponding type and field definitions to ensure consistency.

% A core feature of ROS is its distributed architecture. Specifically, it consists of multiple distributed nodes, each of which is an independent process responsible for executing specific tasks.
% Nodes in ROS interact through three types of interfaces, i.e., topic, service, and action, which are shown in Figure \ref{fig:ROS-node-graph}. The data transmitted through these interfaces is pre-defined and must conform to the corresponding type and field definitions to ensure consistency.
% ROS uses an interface definition language (IDL) to describe these interfaces, and the data transmitted between interfaces must conform to the corresponding definitions.

\textbf{Topic.} Topics are the most prevalent communication mechanism between ROS nodes, which operate based on the asynchronous \textit{publish-subscribe model}.
This model is particularly suitable for handling continuous data streams that demand frequent updates, such as the state of the robot, sensor data, and pose information.
% Topics are suitable for continuous data streams that require frequent updates, such as robot state, sensor data, and pose data.
Nodes can publish messages to a certain topic, and all nodes subscribed to that topic will receive the corresponding messages.
This mechanism is flexible as it allows for multiple publishers to contribute to a single topic, and multiple subscribers can listen to the same topic.
Moreover, each node has the capability to both publish to and subscribe to multiple topics.

\textbf{Service.} Services operate on a \textit{call-and-response model}, consisting of well-defined request-response pairs that allow client nodes to send requests to server nodes.
Upon receiving a request, the server carries out necessary computations and returns the result.
While multiple clients can utilize a single service, only one server can be registered within the same service. Services are ideal for operations that require immediate confirmation and feedback, such as executing commands to ensure correct completion or querying specific data to retrieve results.
Unlike topics, which offer a continuous data stream, services provide data solely in response to calls.
% Each service can be used by multiple clients, but only one server can be registered within the same service.
% Services are suitable for operations that require immediate confirmation and feedback, such as executing commands to ensure correct completion or querying specific data to retrieve results.
% Compared to topics that can provide a continuous data stream, services only provide data when they are called.

\textbf{Action.} An action consists of three parts: goal service, feedback topic, and result service.
The action client initiates a task by sending a goal request to the action server, which decides whether to accept or reject the goal and notify the client via a goal response.
Once the goal is accepted, the client can receive continuous updates on the task execution progress through the feedback topic.
The goal service allows the client to cancel a goal midway or send a new goal to preempt it.
The action server manages each task by maintaining a state machine for each accepted goal, which tracks states such as executing, canceling, and succeeded \cite{ROS2Design-Goal-states}.
Upon task completion or cancellation, the action server communicates the result to the action client through the result service.
Actions are suitable for handling tasks that take a long time to complete and require tracking of the execution progress, such as navigating to a specific point.

In addition to the three interfaces, another crucial mechanism for interaction between ROS nodes is the \textit{callback function}.
The callback function allows ROS nodes to process incoming data transmitted through the above interactions, i.e. topics, services, and actions. When a node subscribes to a topic, registers a service, or initiates an action, it often specifies a callback function that will be invoked whenever a new message, request, or response is received. 
% Callback functions perform appropriate operations based on the received data and the type of interaction, such as executing service requests and adding the results to the response.  
The scheduling and execution of callback functions are managed by the ROS executor \cite{docs-executor} to meet the requirements for real-time performance and responsiveness.

% Callback functions facilitate asynchronous interaction between ROS nodes. A callback function is registered within a specific node for It is invoked when specific data arrives, i.e., messages, requests, or responses, allowing the node to process the incoming data efficiently.
% The \textit{callback function} within the node is invoked to handle the incoming data. % 为了支持以上几种方式的异步通信，ROS采用callback functions来接收数据

% \vspace{-0.1cm}
\subsection{Interaction Bugs in Robotic Systems}

\begin{figure}
    \centering
    \includegraphics[width=0.87\linewidth]{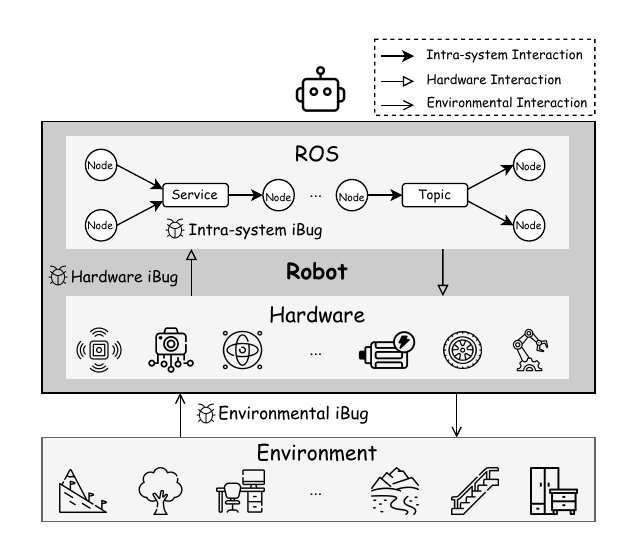}
    \caption{Three categories of \BugNames in robotic systems.}
    \label{fig:InteractionBug}
\end{figure}

Throughout the operations of robots, there exist diverse types of interactions, which refer to the exchange or transfer of information between two or more entities within the system.
For example, a control node may communicate with a motor actuator to execute a specific movement based on the environmental data perceived by the camera.
% These entities can be broadly categorized into three types: \red{nodes} within the system, hardware components of the robot, and external elements in the robot's physical environment.
Such intricate interactions challenge the reliability of robotic systems, as bugs are not confined to individual components, but can also emerge from the interplay between them.
% Understanding the characteristics of such \BugNames plays a critical role in the development of resilient and efficient robotic systems.
\BugNames in robotic systems inherently involve complex interaction mechanism in ROS, heterogeneous hardware coordination, and dynamic physical environments, amplifying challenges beyond traditional distributed systems.
In this work, \BugNames are classified into three categories, i.e., \textit{intra-system \BugNames}, \textit{hardware \BugNames}, and \textit{environmental \BugNames}, as illustrated in Figure~\ref{fig:InteractionBug}.

% In this work, we study three categories of interaction bugs, i.e., \textit{intra-system \BugNames}, \textit{hardware \BugNames}, and \textit{environmental \BugNames}, as illustrated in Figure~\ref{fig:InteractionBug}.

% These bugs represent the full spectrum of interaction scenarios during the operations of robotic systems.
% , from hardware components to software applications and their interface with the environment.
% Specifically, intra-system \BugNames are defects related to \red{node interactions} within the system, hardware \BugNames involve malfunctions or miscommunications with physical components, and environmental \BugNames arise from unexpected external conditions affecting the robot's correctness or performance.

% Robotic software faces a large number of interactions during operation due to its modular nature, the integration demands of software and hardware, and the dynamic and complex physical environment. This presents developers with challenges in accurately crafting software to manage these interactions. In our study, we focus on three types of interaction bugs (\BugNames) in robotic software that can significantly impact the reliability of robots, including \softwareinteraction~bugs, \hardwareinteraction~bugs, and \environmentinteraction~bugs.

\textbf{\Softwareinteraction~bugs.} The flexible communication infrastructure of ROS enables nodes to interact efficiently and effectively through customizable data structures, facilitating various robotic applications.
% Robotic systems often employ a modular design, where different nodes are engaged in extensive interactions to function effectively.
% This modularity, while beneficial for development and scalability, introduces complexity in the form of intra-system interaction bugs.
% Intra-system interactions refer to the exchanges and communications that occur between the nodes within the robotic systems. 
However, bugs can manifest when there is a failure or mismanagement in these internal communications, resulting in malfunctioning or incorrect outcomes.
For example, a ROS node might not receive data if there is a typographical error in the topic name during the subscription process, or an action server node might consistently reject all incoming goal requests due to a mismanaged state from a previous goal.
% \zb{physical collision examples?} \blue{Physical collisions can also occur, such as when a user request triggers a crash in the receiving control node, resulting in the robot moving dangerously even after colliding with an obstacle.}
These bugs highlight the importance of precise interaction configuration and management within the system to ensure robust communications. 
% These bugs highlight the importance of precise configuration and error handling within the system to ensure robust communications. 

% Robotic software typically employs a modular design, where modules must engage in extensive interactions to achieve collaboration. \Softwareinteraction~is the interaction between basic modules within the robotic software, i.e., the interaction between nodes in ROS. \Softwareinteraction~bugs arise from the improper handling of \softwareinteraction, leading to interaction failures or incorrect interaction outcomes. For instance, a ROS node may fail to receive data due to a typo in the topic name during topic subscription, or an action server node may reject all incoming goal requests due to incorrect handling of the previous goal state.
% 
% 这段可以挪去intro Robotic software typically employs a modular design and each module only implements a specific function.
% as it must integrate a variety of complex functions, such as perception, planning, and control.
% The interaction among software modules, i.e., \softwareinteraction, is necessary for them to collaborate. %在我们的研究对象ROS中，具体refers to interaction between nodes

\textbf{\Hardwareinteraction~bugs.} In robotic systems, the software component communicates with the physical hardware by transmitting commands to the hardware, and in turn, receiving data from it.
Hardware interaction bugs arise from flawed management of these software-hardware exchanges, which can lead to interaction breakdowns or unexpected behaviors.
These issues can manifest as an inability to receive data from the hardware or improper execution of commands by the physical components.
For example, when a preexisting control code is deployed without necessary adaptations on new hardware, the robot's actions may deviate from expected behavior due to hardware heterogeneity.

% Therefore, it becomes imperative to ensure the seamless integration of software directives with diverse hardware functionalities to mitigate the occurrence of such interaction bugs.

% Robotic software interacts with physical hardware by sending commands to and receiving data from it.
% %through specific communication interfaces like serial ports. 
% \Hardwareinteraction~bugs are caused by improper handling of interactions between software and hardware, which can lead to interaction failures or incorrect results, such as the inability to receive data from or commands being executed by the physical hardware incorrectly. For instance, when the same control code is not adapted for new hardware, the actions performed by the robot may not align with expectations due to hardware heterogeneity. 
% the robotic software stops running while attempting to zero the sensor due to an incorrect interface configuration for hardware interaction, 
% The interaction with hardware is crucial for robotic software to monitor and control physical robots.
% Two major factors are causing \hardwareinteraction~ bugs: 1) Hardware heterogeneity. When the same code is not adapted for new hardware, the robot's actions may not align with expectations; and 2) The complexity of hardware interfaces. Robotic software often abstracts the use of low-level communication interfaces into hardware interfaces. This process is complex and any mistake can lead to \hardwareinteraction~bugs.

% \vspace{-0.6cm}
\textbf{\Environmentinteraction~bugs.}
Robotic systems interact with the environment by perceiving changes in the surroundings (through sensor data interpretation), making decisions via planning, and adjusting behaviors through control.
This allows the robot to perform tasks effectively in dynamic and complex environments.
% Robotic software interacts with the environment by perceiving changes in the surroundings through interpreting the data received from sensors, making decisions through planning, and adjusting the robot's behavior through control, to enable the robot to perform tasks in a dynamic and complex environment.
Environmental \BugNames arise when there is a mismatch between the outcomes of robotic systems and the actual conditions of the environment based on specifications.
They are typically caused by unexpected external conditions or internal flaws in relevant steps, primarily affecting the robot's ability to accurately perceive its environment or respond appropriately to external changes.
For example, due to a flaw in the task planning implementation, a robot's path is calculated only at the beginning of a task, rather than continuously as expected.
This results in the robot hesitating and retreating for a long time when new obstacles are placed in the environment, instead of promptly recalculating its path to avoid them.
Such environmental interaction bugs often cause the robot to respond to changes improperly and fail to complete the desired tasks.
% This results in the robot hesitating and retreating for extended periods when new obstacles appear, instead of promptly recalculating its path to avoid them. 
% a robot's path is only calculated at the beginning of the task rather than every second as expected due to the incorrect algorithm. This causes the robot to hesitate and retreat for a long time when new obstacles are placed in the environment, rather than avoid them in time through a newly calculated path.
% , which can be caused by factors such as misinterpretation of sensor data, flawed algorithms, and incorrect control logic. 
% \Environmentinteraction~bugs can typically cause the robot to improperly respond to the environment and perform tasks, leading to unexpected behavior.
\section{Methodology}
\label{sec:methodology}

This section outlines the methodology of our empirical study. 
To address the three research questions, we conduct a systematic investigation of \BugNumber \BugNames. 
% A systematic investigation of \BugNumber \BugNames is conducted to address the three research questions.
The following introduces how we choose the target projects (Section \ref{subsec:target projects}), how we collect (Section \ref{subsec:bug collection}) and analyze (Section \ref{subsec:bug analysis}) \BugNames, and the threats to validity (Section \ref{subsec:threats to validity}).

\subsection{Target Projects}
\label{subsec:target projects}

\begin{table}
\centering
\caption{Target projects in our study}
\label{tab:projects}
 \resizebox{\linewidth}{!}
 {
\begin{tabular}{cccccl} 
\specialrule{0.35mm}{0em}{0em}
\textbf{Project} & \textbf{Stars} & \begin{tabular}[c]{@{}c@{}}\textbf{\#Issue\&PR}\\\textbf{Closed}\end{tabular} & \begin{tabular}[c]{@{}c@{}}\textbf{\#Bug}\\\textbf{Closed}\end{tabular} & \begin{tabular}[c]{@{}c@{}}\textbf{\#iBug}\\\textbf{Collected}\end{tabular} & \multicolumn{1}{c}{\textbf{Description}} \\ 
\specialrule{0.15mm}{0em}{0em}
\specialrule{0.15mm}{.1em}{0em}
Navigation2          & 2.7k & 4731 & 164 & 33 & \begin{tabular}[c]{@{}l@{}}Flexible, scalable navigation framework \\for autonomous robots.\end{tabular} \\ 
\hline
realsense2\_camera   & 2.7k & 3144 & 28  & 2  & \begin{tabular}[c]{@{}l@{}}ROS drivers for Intel RealSense cameras, \\enabling 3D depth sensing.\end{tabular} \\ 
\hline
MoveIt2              & 1.2k & 2874 & 455 & 23 & \begin{tabular}[c]{@{}l@{}}Motion planning framework for robotic arms,\\enabling complex manipulations.\end{tabular} \\ 
\hline
MAVROS               & 924  & 1625 & 135 & 15 & \begin{tabular}[c]{@{}l@{}}Communication interface between ROS and\\MAVLink autopilots, used in drones.\end{tabular} \\ 
\hline
ros2\_control~       & 535  & 1800 & 118 & 19 & \begin{tabular}[c]{@{}l@{}}Modular framework in ROS for managing \\robot hardware and controllers.\end{tabular} \\ 
\hline
\begin{tabular}[c]{@{}c@{}}Universal\_Robots\\\_ROS2\_Driver\end{tabular} & 456  & 1102 & 28  & 4  & \begin{tabular}[c]{@{}l@{}}ROS interface for controlling\\Universal Robots' manipulators.\end{tabular} \\ 
\hline
ros2\_controllers    & 403  & 1310 & 74  & 7  & \begin{tabular}[c]{@{}l@{}}Collection of controller interfaces and \\implementations for precise robot \\hardware control in ROS.\end{tabular} \\ 
\hline
depthai-ros          & 250  & 504  & 93  & 4  & \begin{tabular}[c]{@{}l@{}}Integration of DepthAI with ROS,\\enabling spatial AI and computer vision.\end{tabular} \\ 
\hline
aerostack2           & 163  & 657  & 125 & 8  & \begin{tabular}[c]{@{}l@{}}Framework for developing autonomous \\aerial robotic systems.\end{tabular} \\ 
\hline
turtlebot4           & 110  & 449  & 182 & 6  & \begin{tabular}[c]{@{}l@{}}Open-source platform for education \\and research in ROS and robotics.\end{tabular} \\
\specialrule{0.35mm}{0em}{0em}
\end{tabular}
}
\end{table}
 
We select ten representative ROS-based open-source robotic software projects for our empirical study, with their statistics shown in Table \ref{tab:projects}. The ``\#BugClosed'' column indicates the number of closed issues and pull requests (PRs) that are labeled as bugs in each project.
% In the table, the ``\#Issue\&PR Closed'' column represents the number of issues and pull requests (prs) that are in a closed state in each project. The ``\#Bug Closed'' column indicates the number of closed issues and prs that are labeled as bugs. The ``\#iBug Collected'' column denotes the number of \BugNames collected from each project.
The projects are chosen through the following steps: 
1) we crawl all the projects with an active status from the official ROS index \cite{ROS-Index} for the Humble distribution, which is currently the most prevalent ROS version \cite{2023-ROS-Metrics-Report}. 
%, and obtain 684 projects.
Next, 2) we exclude projects not located at the application layer, including middleware for inter-process communication and internal ROS client libraries such as \textit{ecal} and \textit{rclcpp}. Additionally, we remove tools designed to assist developers in development and debugging such as \textit{ros2cli} and \textit{rosbag2}, and standalone C++ computation libraries not based on ROS like \textit{Pinocchio} and \textit{OMPL}. After this step, we obtain 184 projects.
% \zx{184 is based on deepseek. if we remove too many projects from ROS index, will we be suspected?}\zb{maybe we just don't mention 684, it's useless to us compared to 184. You can say there are hundreds of row projects, but do not mention 684}
Then, 3) we crawl each GitHub repository to gather its number of stars and issues (including PRs) with a closed status, as they are more likely to reach definite conclusions. For non-GitHub projects, we manually open their links to analyze their statistics.
Finally, 4) we narrow our focus to projects with more than 100 stars and 400 issues, resulting in the following: navigation2, MoveIt2, MAVROS, aerostack2, Universal\_Robots\_ROS2\_Driver, depthai-ros, realsense2\_camera, ros2\_control, ros2\_controllers, and turtlebot4. 
% 用Issue number来筛选项目是因为这个表明了项目的popularity

\subsection{Bug Collection}
\label{subsec:bug collection}
For each project, we focus on issues and pull requests that have both a closed status and a bug label, as they are more likely to reach definite conclusions and to be actual bugs than others. In total, the selected projects contain 1,402 such bugs.
% For each project, we focus on issues and prs that are closed and labeled as bugs, as they are more likely to have definite conclusions and to be actual bugs than others. The target projects contain a total of 1,402 bugs, i.e., issues and prs with both a closed status and a bug label.
To collect \BugNames, we use related keywords such as ``topic'', ``callback'', ``hardware'',  ``behaviour'', and ``interact'' to search for potential \BugNames. Then, we manually confirm each bug by examining the bug description, discussion, and fix patch. The keywords are iteratively refined in this process. 
We exclude the bugs that are duplications of already collected bugs, lack a clear description, have no available fix patch, or can not be fully understood. This step returns \FullBugNumber \BugNames in total. 
We randomly sample one-third of these and ultimately select 121 \BugNames for further study.
% We randomly sample from these \BugNames and finally screen out \BugNumber \BugNames~ for further study. 
% 364/1402=0.26 ，然后再随机抽取三分之一（相当于所有Bug的四分之一是iBug，然后再抽了三分之一）

% exclude the bugs that are duplications of already collected bugs, lack a clear description, have no available fix patch, or can not be fully understood

\subsection{Bug Analysis}
\label{subsec:bug analysis}
% opencoding + saturation point
% 每一轮单独标记，完成后讨论并解决冲突，直到标签saturation，没有新的label产生
    
We perform an in-depth analysis of the collected bugs following an open coding procedure \cite{DBLP:journals/tse/Seaman99-OpenCodingProcedure}. 
Three authors with extensive experience in ROS manually analyze the bugs and label their interactions, root causes, fixes, and impacts. For each bug, we investigate the bug description, discussion, fix patch, and log traces (if available) to extract enough information to label its category of interaction and document a detailed account of the bug. For the root causes, we locate the bugs based on the extracted information to determine the code-level causes. For bug fixes, we carefully review the fix patches in the corresponding pull requests as well as other relevant information and summarize common fix patterns addressing the same root causes. The bug impacts caused by \BugNames~are determined from the bug reports along with other useful extracted information.

The bug labeling process is performed iteratively. In each iteration, each author individually labels a different subset of bugs that they had not labeled in the previous iterations. The authors first attempt to label the bugs using existing categories and only create a new label when a suitable category cannot be found. At the end of each iteration, the authors discuss and resolve all their problems and conflicts. After three iterations, a saturation point is reached where the taxonomy remains stable without changes to the labels.

\subsection{Threats to Validity}
\label{subsec:threats to validity}
% In this section, we discuss potential threats to the validity of our empirical study.

\textbf{Internal validity.} Threats to internal validity relate to the manual analysis of bugs. 
To minimize biases introduced by subjective judgment, each bug was reviewed by three authors. Any problems or classification conflicts were resolved through extensive discussion until a consensus was reached.
All content within each bug, including the bug report, buggy code, and patches, was thoroughly examined. 
% To minimize biases introduced by subjective judgment, we thoroughly examined all the content within the bug reports, as well as the buggy code and patches. Each bug was reviewed by three authors, and any problems or classification conflicts were resolved through extensive discussion until a consensus was reached.
Bugs without a clear description or could not be fully understood were excluded. 
Thus, the bugs in our study have been comprehensively analyzed.
% Thus, the bugs in our study are \red{believed to be valid} and have been comprehensively analyzed.

\textbf{External validity.} Threats to external validity relate to the representativeness and generalizability of our study. 
To ease such threats, we chose ROS as the target system for our empirical study, which is the most popular framework and the \textit{de facto} standard for robotic software development. We crawled all the active projects from the official ROS index and carefully selected the targets, covering a variety of robotic applications with different functionalities in robotic systems. 
Therefore, the projects and bugs we selected should have representativeness and the findings in our study can be generalized to the projects that we have not studied.

% The modular design similar to ROS is shared by other robotic systems, such as YARP \cite{metta2006yarp}, Orca \cite{DBLP:conf/icra/BrooksKMWO05-Orca}, and Player \cite{DBLP:conf/iros/GerkeyVSHSM01-player}, where intricate interactions also exist. 

\section{Study Results}
\label{sec:study_results}

\subsection{Root Causes}
\label{sec:root_cause}

\begin{figure*}
    \centering
    \includegraphics[width=\textwidth]{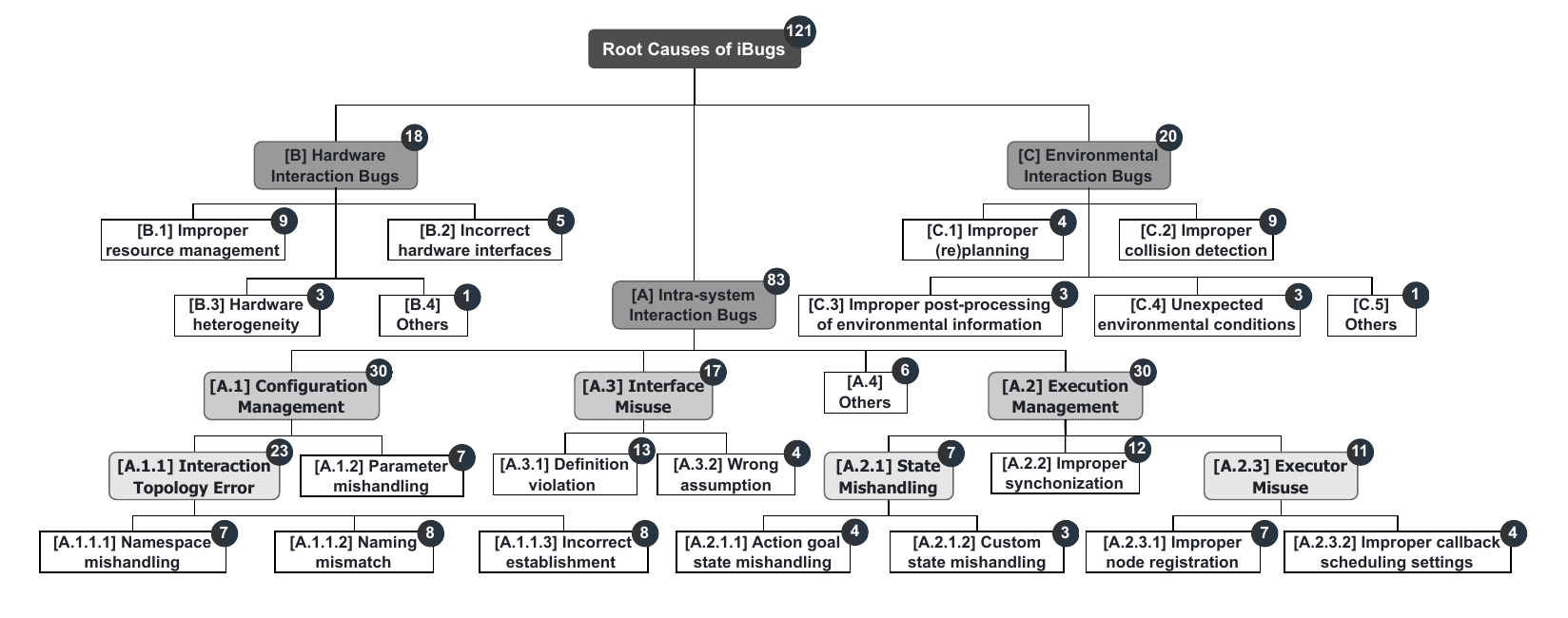}
    \caption{The taxonomy of root causes for \BugNames.}
    \label{fig:root_cause}
\end{figure*}

We investigate how \BugNames are introduced by developers and manually classify their root causes. The taxonomy of root causes is shown in Figure \ref{fig:root_cause}. Each rounded rectangle with either a dark or light gray background represents a subcategory and each rectangle with a white background denotes an atomic category. The number in the circle attached to each rectangle indicates the number of \BugNames in the corresponding category. Each \BugName is classified into only one atomic category.

% ROS中一个核心的特性是，存在大量的message在节点和topic之间流动以实现ROS内部各个节点之间的交互。由于节点之间的差异性，开发者可能会对不熟悉的节点作出错误假设，或者遗漏交互过程中的某些重要步骤，导致出现各种iBugs.
% 这一章节的Findings：最打的部分是缺乏线程间同步和消息的不正确处理。然而，namespace topic/servername 这两个涉及人工配置的也容易导致各种Bug，值得留意
\subsubsection{\SoftwareInteraction~Bugs}
This category contains 83 \BugNames arising from \softwareinteraction s. Developers can use mechanisms provided by ROS such as topic, service, and action to enable the \softwareinteraction~between distributed ROS nodes. Developers may mistakenly handle certain steps during the interaction or make wrong assumptions about unfamiliar nodes due to the discrepancies between various nodes, resulting in \BugNames that belong to this category.

\begin{figure}[H]
    \centering
    \includegraphics[width=\linewidth]{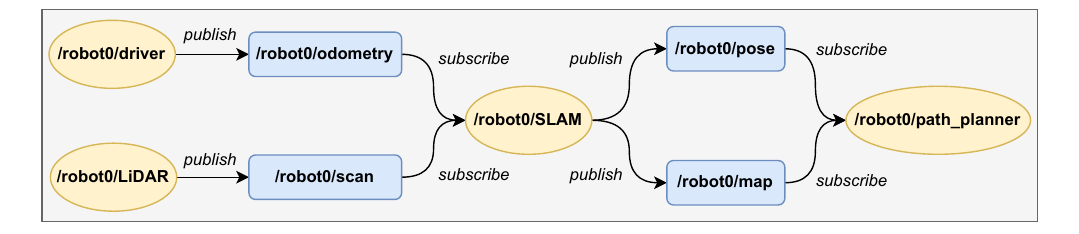}
    \caption{An illustrating example of ROS interaction topology. Yellow ovals represent nodes and blue rounded rectangles represent topics.}
    \label{fig:topology_example}
\end{figure}

\textbf{Configuration Management [A.1] (30 bugs).} Configuration management is crucial to coordinate various components in robotic systems. Developers bear the responsibility of configuring the organization and interconnection of various ROS nodes, 
where any mismanagement can lead to intra-system \BugNames.
% each playing a vital role in facilitating the \softwareinteraction s. 

    \textit{Interaction Topology Error [A.1.1] (23 bugs).} Interaction topology refers to the structure of how different ROS nodes are organized and interact with each other in a robotic system.
    % , which can be directly visualized using tools such as rqt \cite{Rqt}. 
    Developers should perform proper configurations to ensure that the node/interface settings and the connection relationships align with expectations; otherwise, misconfigurations can lead to \BugNames. 
    
    We explain the three atomic categories through an example from Robucar~\cite{DBLP:journals/evi/Robucar}, which is shown in Figure \ref{fig:topology_example}. 
    % The original figure has been modified, and only nodes and topics are reserved for better illustration. 
    The original figure has been modified, retaining only nodes and topics for better illustration. 
    % The shared prefix ``robot0'' is the namespace that represents this robot. 
    % The ``/robot0/driver'' and ``/robot0/LiDAR'' nodes publish odometry (tracks a robot's movement using wheel or sensor measurements) and laser scan data to ``/robot0/odometry'' and ``/robot0/scan'' topics, respectively. 
    The ``/robot0/driver'' and ``/robot0/LiDAR'' nodes publish odometry and laser scan data to ``/robot0/odometry'' and ``/robot0/scan'' topics, respectively. 
    The ``/robot0/SLAM'' node subscribes to these topics, processes the data, and publishes the estimated pose and generated map. The rest of the figure follows a similar pattern.
    
    % The ``/robot0/SLAM'' node subscribes to both topics and processes the received data to generate a map of the environment and estimates the robot's pose. Subsequently, it publishes the computed data to topics ``/robot0/map'' and ``/robot0/pose''. 
    % Both topics are subscribed to and the data is used by the node ``/robot0/path\_planner'' to calculate the path for the robot to navigate within the environment.
    % processes the data to create a map of the environment and estimates the robot's pose.
    % The node ``/robot0/SLAM'' subscribes to and receives data from both topics. Using this information, it processes the data to create a map of the environment and estimates the robot's pose, which it then publishes to the /robot0/pose'' and /robot0/map'' topics.
    % generates a map of the environment and estimates the robot's pose based on the data it receives.
% 开发人员应正确配置节点交互拓扑，使得接口设置与数据流向符合预期，否则会带来iBug.
    % Bugs arising from errors in the interaction topology, such as connection or establishment faults for nodes/interfaces, are categorized under this classification.

    \begin{itemize}[noitemsep,leftmargin=5.5mm]
        \item  \textit{Namespace mishandling [A.1.1.1] (7 bugs).} The namespace groups relevant nodes and interfaces in ROS to avoid naming conflicts. For example, the shared prefix ``robot0'' in Figure \ref{fig:topology_example} groups all nodes and topics associated with the same robot.
        % ensuring that different robots within the system remain independent and do not interfere with each other.
        % When another similar robot needs to operate within the same system, it can be placed under the namespace ``robot1''. This ensures that the interactions between the two robots remain isolated and do not interfere with each other.         % It helps manage large robotic systems by allowing different components to have similar names while they are still uniquely identified within their namespaces.
        However, the namespace must be manually handled in various locations, including code and configuration files, and thus becomes error-prone. 
        % For instance, developers may declare the namespace in launch files but neglect to propagate it in the code, causing the node to fail in finding correct interaction targets (\href{https://github.com/moveit/moveit2/issues/2415}{MoveIt2\#2415}). 
        For instance, if the SLAM node in Figure \ref{fig:topology_example} fails to specify its designated namespace, ``robot0'', in its launch file, all interactions that depend on it will be disrupted.
        % Consider the SLAM node in Figure \ref{fig:topology_example}: if it forgets to specify the namespace ``robot0'' in its topic subscription, it will fail to receive data from the topics. % 或者这类也可以用suppose...
        % 这个例子有点像naming mismatch, 决定换个例子
        % For instance, a node's configuration file may incorrectly specify the namespace of the topic it intends to subscribe to (\href{https://github.com/turtlebot/turtlebot4/issues/247}{Turtlebot4\#247}). Consider the SLAM node in Figure \ref{fig:topology_example}: if it mistakenly assigns the namespace of the "/robot0/scan" topic as "myrobot", it will fail to receive laser scan data from the topic in such a case. % 或者这类也可以用suppose..    
        % For example, the node handler in \href{https://github.com/mavlink/mavros/issues/207}{MAVROS\#207} has declared the namespace ``visualization'' during initialization, and the developer mistakenly includes the namespace again when retrieving parameters from the server, causing the system to incorrectly look for parameters under the ``visualization/visualization'' namespace, which leads to failure in retrieving parameters. 
    
        \item \textit{Naming mismatch [A.1.1.2] (8 bugs).} The naming of the interaction interfaces must be manually specified by developers.        
        Developers may assign inconsistent names, thus causing \BugNames due to naming mismatches.
        % Developers may assign inconsistent names in the interaction, leading to \BugNames due to naming mismatches.  
        Assuming that the LiDAR node in Figure \ref{fig:topology_example} mistakenly publishes data to the  ``/robot0/laser\_scan'' topic but the SLAM node remains unchanged, such a mismatch will prevent the SLAM node from receiving the expected data.    
        % In Figure \ref{fig:topology_example}, the LiDAR node publishes data to the topic ``/robot0/scan''. Assuming that the SLAM node mistakenly subscribes to the topic ``/robot0/laser\_scan'', this mismatch will prevent the node from receiving the expected data, as it will have subscribed to a non-existent topic.
        % Suppose the SLAM node in Figure \ref{fig:topology_example} erroneously specifies the topic name as ``/robot0/laser\_scan'' instead of the intended ``/robot0/scan''. In that case, the node will fail to receive LiDAR data from the expected source.
        
        % The naming of the interfaces and nodes requires manual configuration by developers. 
        % Some of the information in \softwareinteraction~is required to be manually configured by the developers in ROS, such as the target topic names for publishing or subscribing data, and the server names for sending service requests or action goals.
        % \Softwareinteraction~bugs occur when there is a mismatch between the expected names of the interaction target in the ROS system. 
        % For example, in \href{https://github.com/ros-navigation/navigation2/issues/1450}{Navigation2\#1450}, the action client declares the action server name to be ``navigate\_to\_pose'' while the server side declares it as ``NavigateToPose''. Such a mismatch causes the goal request to be ignored since the server does not receive that.
        \item  \textit{Incorrect establishment [A.1.1.3] (8 bugs).} This category concerns the incorrect establishment of nodes and interfaces in the interaction topology. 
        % For example, developers may establish two services with the same name (\href{https://github.com/moveit/moveit2/issues/2731}{MoveIt2\#2731}), forget to establish a publisher (\href{https://github.com/turtlebot/turtlebot4/issues/103}{Turtlebot4\#103}), or retain a deprecated and redundant publisher. 
        Assuming that the publisher in the LiDAR node is not created in Figure \ref{fig:topology_example}, its data will not be published to the topic, thereby affecting subsequent interactions.
    \end{itemize}

\begin{tcolorbox}[colframe=black!75!white, colback=white!90!gray, boxrule=0.2mm, arc=1mm, boxsep=0.5mm, width=\linewidth]
\refstepcounter{finding}
\textbf{Finding \thefinding}: \textit{The interaction topology error is a common source of \BugNames, accounting for 19.01\% (23/121) of the total \BugNames.}
\label{finding:root_cause-topology_error}
\end{tcolorbox}

    \textit{Parameter mishandling [A.1.2] (7 bugs).} 
    In ROS, parameters store data shared by multiple nodes and reside on a specific node (the parameter server). 
    They can be preconfigured and are dynamically accessible at runtime through services for reading and modification. 
    % Parameters are used to store data commonly used by multiple nodes in ROS. They are stored on a specific node (namely the parameter server) and can be preconfigured. Parameters are dynamically accessible by other nodes through services for reading and modification during runtime. 
    Developers may incorrectly implement the access logic for parameters, leading to \BugNames such as ineffective parameter updates or inconsistencies between retrieved and pre-configured values. 
    % Mismanagement in these parameters can result in \BugNames, e.g., the parameter values are updated but have no effect, and the retrieved parameter values are inconsistent with the pre-configured ones. 
    For instance, in \href{https://github.com/ros-navigation/navigation2/issues/2686}{Navigation2\#2686}, although the downsampling option is set to false, the downsampling factor is still being applied in the calculation of the minimum turning radius. The robot's motion is affected since the turning radius is incorrectly reduced.
    % #67 有时间可以画个图说明，比较有意思

\textbf{Execution Management [A.2] (30 bugs).} In the interactive execution of robotic systems, it is essential to properly manage the dynamic characteristics of the system, including state transitions, synchronization, and the utilization of the executor; otherwise, \BugNames can occur.

    \textit{State Mishandling [A.2.1] (7 bugs).} 
    % Developers must properly maintain a state to monitor the robot's task execution progress and ensure the tasks are carried out correctly. 
    State transitions often occur during the \softwareinteraction~in order to monitor the robot's task execution progress and ensure the tasks are properly carried out. For example, the goal state may transition to ``ACCEPTED'' when the client requests the action server to start an action execution. 
    The states we mention here include the ROS built-in states, i.e., the goal states in the Action mechanism \cite{ROS2Design-Goal-states}, and user-defined states such as whether the camera is running or not. 
    
    Developers may not clearly understand when or how to inspect or transition between states and cause \BugNames. 
    For example, in \href{https://github.com/luxonis/depthai-ros/issues/238}{Depthai-ROS\#238}, the camera’s current state is not checked in the callback function that handles the camera startup request. As a result, a duplicate attempt to start the camera causes the software to crash. In another bug \href{https://github.com/ros-navigation/navigation2/pull/918}{Navigation2\#918}, a crash occurs due to an incorrect goal state transition that is not defined in the state machine.
    % an incorrect goal state transition which is undefined in the state machine also causes a crash. 
    % 
    Four bugs resulting from mishandling of goal states were categorized under \ul{\textit{[A.2.1.1] Action goal state mishandling}}, while three bugs arising from mishandling of user-defined states were classified under \ul{\textit{[A.2.1.2] Custom state mishandling}}.

    \textit{Improper synchronization [A.2.2] (12 bugs).} \BugNames in this category arise when threads on the same or different nodes are not properly synchronized. They are primarily induced by the specific sequence in which callbacks are executed or the precise timing of interaction occurrences.
    Take \href{https://github.com/ros-navigation/navigation2/issues/2931}{Navigation2\#2931} as an example, the ACML node frees the old map's memory in the callback handling a newly received map message. Meanwhile, a laser scan message arrives at the same node and triggers another callback, which accesses the freed map and causes a segmentation fault. 
    % In \href{https://github.com/UniversalRobots/Universal_Robots_ROS2_Driver/issues/148}{Universal\_Robots\_ROS2\_Driver\#148}, a crash occurs when the service from a controller node is requested by another node before the controller is activated.      
    
    %讲#25（两个回调同时访问map带来崩溃）、#11（服务在节点激活前就被调用了）
    % when the set\_io service from gpio\_controller node is called by another node before the gpio\_controller node is activated, the gpio\_controller node will crash. 

    \textit{Executor Misuse [A.2.3] (11 bugs)} Callback functions within ROS nodes are scheduled by the executor \cite{docs-executor}. Each node must specify its scheduling rules (reentrant or mutually exclusive) for the respective callback functions, and these functions are only executed after both the node has been registered with the executor and the executor has started spinning. Developers may inadvertently misuse the executor, thereby impeding the proper execution of interactions.

\begin{itemize}[noitemsep,leftmargin=5.5mm]
        \item \textit{Improper node registration [A.2.3.1] (7 bugs).} 
        % This category is related to developers improperly registering nodes with the executor. Developers may either omit, duplicate registrations, or employ incorrect registration methods.
        This category involves incorrect node registration with the executor, including omissions, duplicates, or improper methods.
        For example, the node in \href{https://github.com/ros-navigation/navigation2/pull/808}{Navigation2\#808} is not registered with the executor, failing to receive incoming messages via the callback function.
        The bug \href{https://github.com/moveit/moveit2/issues/1127}{MoveIt2\#1127} registers the same node twice with the executor, causing a crash. 
        In \href{https://github.com/ros-navigation/navigation2/issues/792}{Navigation2\#792}, the node was continuously registered to and removed from the executor within a loop, resulting in significant CPU consumption.
        
        \item \textit{Improper callback scheduling settings [A.2.3.2] (4 bugs.)} The callback scheduling policies in ROS are determined by the combination of executors (single/multi-threaded) and callback groups (reentrant/mutually exclusive). Improper configurations may lead to deadlocks or resource conflicts. 
        % ROS suports both single- and multi-threaded executors, which can be used in conjunction with reentrant or mutually exclusive callback groups. These groups define whether callback functions within the same group can execute concurrently. The combination of these configurations determines the scheduling policy of callback functions. Improper configurations may lead to deadlocks or resource conflicts.      
        % ROS provides both single-threaded and multi-threaded executors, which can be used in conjunction with reentrant or mutually exclusive callback groups. These groups define whether callback functions within the same group can execute concurrently. The combination of these configurations determines the scheduling policy of callback functions. Improper configurations may lead to deadlocks or resource conflicts. 
        Take \href{https://github.com/mavlink/mavros/issues/1657}{MAVROS\#1657} as an example, the timer callback  must wait for the future object's done-callback to complete before proceeding. However, since both belong to the same mutually exclusive callback group, they are never scheduled to run simultaneously, causing a deadlock.
        % Take \href{https://github.com/mavlink/mavros/issues/1657}{MAVROS\#1657} as an example, the timer callback needs to wait for the future object's done-callback, which handles the incoming service response, to complete before it can proceed. However, both callbacks belong to the same callback group, preventing them from executing concurrently, ultimately leading to a deadlock.
    \end{itemize}
    
\begin{tcolorbox}[colframe=black!75!white, colback=white!90!gray, boxrule=0.2mm, arc=1mm, boxsep=0.5mm, width=\linewidth]
\refstepcounter{finding}
\textbf{Finding \thefinding}: \textit{9.10\% (11/121) of the iBugs originate from the misuse of the ROS executor.}
\label{finding:root_cause-executor_misuse}
\end{tcolorbox}

\textbf{Interface Misuse [A.3] (17 bugs).} The three types of interfaces—Topics, Services, and Actions (see Section \ref{subsec:Background-ROS})—serve as the primary mechanisms for \softwareinteraction s in ROS. Within these interfaces, data exchange is facilitated through specific data structures: messages for Topics, request/response pairs for Services, and goal/feedback/result sequences for Actions. They have strict field and type definitions, allowing developer customization in configuration files. 
However, \BugNames can occur if developers deviate from the definition when sending or receiving data with these interfaces.

\begin{itemize}[noitemsep,leftmargin=5.5mm]
    \item \textit{Definition violation [A.3.1] (13 bugs).} 
    \BugNames in this category arise when senders violate data structure definitions. Developers may inadvertently omit field assignments or assign values that do not adhere to the format or content specifications. 
    % \BugNames in this category are caused by the sender's data construction violating the definitions specified in the configuration files. Developers may inadvertently omit field assignments or assign values that do not adhere to the format or content specifications. 
    For example, the y-size field of a bounding box is not assigned in \href{https://github.com/luxonis/depthai-ros/issues/258}{Depthai-ROS\#258}, leaving the subscriber with incomplete data.
    In \href{https://github.com/mavlink/mavros/pull/1852}{MAVROS\#1852}, the publisher does not construct the timestamps in message headers using seconds and nanoseconds, which should be the correct format.
    
    % In \href{https://github.com/mavlink/mavros/pull/1852}{MAVROS\#1852}, the publisher does not correctly construct the timestamps in message headers using seconds and nanoseconds, thereby publishing messages with wrong formats and confusing the subscribers.

    \item \textit{Wrong assumption [A.3.2] (4 bugs).} 
    % This category relates to cases where developers make incorrect assumptions about the data received via ROS interfaces. 
    This category relates to cases where developers make incorrect assumptions about the data they receive through ROS interfaces. 
    % For example, in \href{https://github.com/mavlink/mavros/issues/1842}{MAVROS\#1832}, the developer of a callback function mistakenly assumes the type of the timestamp in the message header , omitting the necessary type conversion. This leads to an invalid comparison between mismatched types, ultimately causing a crash.
    Take \href{https://github.com/mavlink/mavros/issues/1842}{MAVROS\#1832} as an example, the callback function responsible for handling incoming pose messages incorrectly assumes that the timestamp in the header is of type \seqsplit{``rclcpp::Time''} and, consequently, omits the necessary type conversion. However, the actual type of the timestamp is \seqsplit{``builtin\_interfaces::msg::Time''}, leading to a comparison between two differently formatted time representations, ultimately resulting in a crash.        
\end{itemize}

\begin{tcolorbox}[colframe=black!75!white, colback=white!90!gray, boxrule=0.2 mm, arc=1mm, boxsep=0.5mm, width=\linewidth]
\refstepcounter{finding}
\textbf{Finding \thefinding}: \textit{Interface misuse is another common cause that affects 14.05\% (17/121) of the total \BugNames.}
\label{finding:root_cause-interface_misuse}
\end{tcolorbox}

\textbf{Others [A.4] (6 bugs).} The remaining intra-system \BugNames are classified as \textit{others}. The root causes of the bugs in this category are diverse. For example, in \href{https://github.com/ros-navigation/navigation2/issues/3477}{Navigation2\#3477}, the service client does not remove pending requests when the server node does not respond. Memory leak takes place when the failed requests pile up.

\begin{tcolorbox}[colframe=black!75!white, colback=white!90!gray, boxrule=0.2 mm, arc=1mm, boxsep=0.5mm, width=\linewidth]
\refstepcounter{finding}
\textbf{Finding \thefinding}: \textit{Intra-system \BugNames constitute the majority of \BugNames, representing 68.60\% (83/121) of the total cases.}
\label{finding:root_cause-intra_system}
\end{tcolorbox}
% The primary root causes include definition violation, improper synchronization, naming mismatch, and incorrect establishment.}
% configuration management (), execution management, and interface misuse.}
% The primary root causes relate to configuration management (24.79\%$\approx$30/121), execution management (24.79\%$\approx$30/121), and interface misuse (14.05\%$\approx$17/121).

\subsubsection{\HardwareInteraction~Bugs} 
The \hardwareinteraction~is essential for the robotic software to monitor and control physical robots. 
For instance, a ROS node can retrieve the present pose of the end effector, which is a part of a robotic arm, and issue commands to move it towards a designated target position. 
For seamless \hardwareinteraction, developers must ensure effective hardware resource management, precise configuration and implementation of hardware interfaces, and sufficient compatibility between software and hardware. Failure to address these aspects can result in \BugNames.
% For example, a ROS node can read the current pose from the end effector, a component of a robotic arm, and issue commands to move it towards a designated target position. 
% 开发者需要确保对硬件资源管理是恰当的，对硬件接口的实现是正确的，并且软件与硬件之间的适配是足够的，否则交互Bug可能会产生
% However, the hardware devices are heterogeneous and the hardware interfaces require complex manual configuration. These factors can lead to \hardwareinteraction~bugs.

\textbf{Improper resource management [B.1] (9 bugs)} Resource management is primarily responsible for the coordination of hardware resource allocation and release, state monitoring, and life cycle management, aiming to ensure efficient and stable hardware interactions. Faults in these processes may result in \BugNames. 

For example, the same joint has been assigned to multiple controllers in \href{https://github.com/moveit/moveit2/issues/2698}{MoveIt2\#2698}, resulting in a resource conflict that leads to control failure. 
In \href{https://github.com/ros-controls/ros2_control/issues/557}{ROS2\_control\#557}, the initialization is omitted when importing the hardware component, resulting in the inability to perform both read and write operations on the hardware.

\textbf{Incorrect hardware interfaces [B.2] (5 bugs).} The hardware interfaces allow the software to abstract the complexities of hardware interactions, enabling developers to focus on higher-level robotic functionalities. However, incorrectly configured or implemented interfaces can lead to \BugNames. 

Take \href{https://github.com/ros-controls/ros2_controllers/issues/1167}{ROS2\_controllers\#1167} as an example, the position command interface is mistakenly implemented as a velocity command interface in the steering controller, although the documentation suggests that the reference for it is position. This \BugName prevents accurate interactions with the vehicle.

\textbf{Hardware heterogeneity [B.3] (3 bugs).} Robotic systems consist of heterogeneous hardware that varies in functionality, performance, manufacturers, etc.
Developers may reuse the same program without proper adaptation to the hardware in use, owing to the convenient reusability of ROS packages. 

For example, in \href{https://github.com/moveit/moveit2/issues/2165}{MoveIt2\#2165}, the gripper turns left instead of moving horizontally while maintaining its orientation as intended, bacause the joint connecting the arm and the gripper has a different max speed from the other joints. This causes it to quickly reach its limit and block the movement of the remaining joints. The control code does not account for this discrepancy and causes an \BugName.

\textbf{Others [B.4] (1 bug).} The remaining one hardware \BugName belong to this category. In \href{https://github.com/ros-controls/ros2_control/issues/1144}{ROS2\_control\#1144}, the hardware system incorrectly performs constraint checks on the interfaces when the dynamic calculations are disabled, preventing the trajectory from executing on the robot.
% with multiple command interfaces.

\begin{tcolorbox}[colframe=black!75!white, colback=white!90!gray, boxrule=0.2mm, arc=1mm, boxsep=0.5mm, width=\linewidth]
\refstepcounter{finding}
\textbf{Finding \thefinding}: \textit{14.88\% (18/121) of \BugNames are hardware \BugNames. Improper resource management is the most prevalent root cause, accounting for 50\% (9/18) of these cases.}
\label{finding:root_cause-hardware}
\end{tcolorbox}

\subsubsection{\EnvironmentInteraction~Bugs} \label{subsec:root_cause-environmental_iBugs}
The robotic system perceives changes in the environment, makes decisions through planning, and adjusts the robot's behavior via control, enabling effective environmental interaction. 
However, any unforeseen and inadequately addressed environmental context, as well as mishandling certain steps of environmental interaction, can lead to environmental \BugNames. 
% However, any unforeseen and inadequately addressed environmental context can give rise to environmental \BugNames. Furthermore, mishandling certain steps of environmental interaction can also result in \BugNames.

\textbf{Improper (re)planning [C.1] (4 bugs).} 
The robot is required to dynamically adapt its behavior through continuous planning and replanning to effectively address ongoing variations in the constantly changing environment.
However, developers may fail to accurately implement the (re)planning functionality of the robot, thereby resulting in erroneous responses to the external.
For example, when an obstacle suddenly blocks the robot's path in \href{https://github.com/ros-navigation/navigation2/issues/1307}{Navigation2\#1307}, the robot does not replan its path at the expected frequency of 1Hz to avoid the obstacle in time due to an error in the implementation of its behavior tree.
% is suddenly placed in the robot's path during its path execution. 
% However, due to an error in the implementation of its behavior tree, the robot does not replan its path at the expected frequency of 1Hz to avoid the obstacle in time.
% However, the robot does not update its path once per second as expected to avoid the obstacle in time due to an error in the implementation of its decision tree.
Instead, it exhibits a recovery behavior \cite{Recovery-bahavior} and hesitates in front of the obstacle for a long time before eventually computing a new path.
% \footnote{The robot's recovery behavior is to retreat some distance in order to reassess the environment and find a new path, usually performed when it gets into trouble.}

\textbf{Improper collision detection [C.2] (9 bugs).}
Collision detection is employed to identify potential undesirable collisions between robots and their environments, thereby enabling proactive avoidance. 
Implementation flaws (e.g., incorrect collision geometries or miscalculated look-ahead distances) may cause iBugs.
% Misimplementations, such as incorrect collision geometries or miscalculated look-ahead distances, leads to iBugs.
% However, developers may inaccurately implement collision detection, such as incorrectly maintaining collision geometries or miscalculating look-ahead distances, causing environmental \BugNames.
For example, the collision geometry was not integrated into the planning scene for collision detection in \href{https://github.com/moveit/moveit2/issues/2548}{MoveIt2\#2548}, which allows the robot to collide with the obstacles. 
In \href{https://github.com/ros-navigation/navigation2/issues/2670}{Navigation2\#2670}, the robot raises false collision warnings when navigating close to obstacles, as the look-ahead point exceeds the actual reachable point of the robot.

\textbf{Improper post-processing of environmental information [C.3] (3 bugs).}
The robotic system processes the raw environmental information collected by sensors for subsequent steps through operations such as noise reduction and distance calculation. 
Improper implementation of these processes can result in erroneous perception of the environment, leading to environmental \BugNames. For example, distortions occur in \href{https://github.com/IntelRealSense/realsense-ros/issues/363}{Realsense-ROS\#363} when converting camera data into a point cloud due to the use of mismatched camera calibration information. Although the robot is positioned on a flat surface, the resulting point cloud forms a slope.

\textbf{Unexpected environmental conditions [C.4] (3 bugs).} Unexpected environmental conditions, particularly edge cases, that have not been adequately addressed by the robotic system can cause \BugNames. For example, in \href{https://github.com/ros-navigation/navigation2/issues/2781}{Navigation\#2781}, the robot continuously navigates in an environment with dynamic obstacles. When multiple obstacles approach the robot, the planner server crashes. This is attributed to the \textit{raytraceLine} function, which is responsible for clearing obstacles, encountering a scenario where the starting and ending points of a line segment coincide, leading to a crash.

\textbf{Others (C.5) (1 bug).} 
We classify the remaining environmental \BugName into this category. 
In \href{https://github.com/ros-controls/ros2_controllers/issues/280}{ROS2\_controllers\#280}, the \textit{diff\_drive \_controller}'s odometry incorrectly computes angular velocity with an opposite direction to the actual value, causing erroneous pose estimation that may confuse the navigation and control systems.
% In \href{https://github.com/ros-controls/ros2_controllers/issues/280}{ROS2\_controllers\#280}, the angular velocity computed by the odometry of \textit{diff\_drive\_controller} is in the opposite direction of the actual value. This discrepancy results in the system's inability to accurately calculate the robot's position and orientation, leading to confusion and errors within the navigation and control systems.

% In \href{https://github.com/ros-controls/ros2_controllers/issues/280}{ROS2\_controllers\#280}, the angular velocity computed by the odometry of \textit{diff\_drive\_controller} is in the opposite direction of the actual value. This discrepancy results in the system's inability to accurately calculate the robot's position and orientation, leading to confusion and errors within the navigation and control systems
% For example, in \href{https://github.com/ros-navigation/navigation2/issues/3361}{Navigation2\#3361}, the zero-velocity command is not published when the robot encounters an unexpected situation due to an error in the exception handling code. As a result, the robot does not stop moving and continues to operate dangerously based on the last velocity command.

\begin{tcolorbox}[colframe=black!75!white, colback=white!90!gray, boxrule=0.2mm, arc=1mm, boxsep=0.5mm, width=\linewidth]
\refstepcounter{finding}
\textbf{Finding \thefinding}: \textit{16.53\% (20/121) of the \BugNames~are environmental \BugNames.
Among these, improper collision detection (45\%$=$9/20) is the most prevalent root cause.}
\label{finding:root_cause-environment}
\end{tcolorbox}

% \textbf{Finding \thefinding}: \textit{16.53\% (20/121) of the \BugNames~are environmental \BugNames.
% Among these, improper collision detection (45\%$=$9/20) and improper (re)planning (20\%$=$4/20) are the two most prevalent root causes.}

\subsection{Fix Patterns}
\label{sec:fix}
% \hline
    % \specialrule{0.35mm}{0em}{0em}
% \hline
    % \specialrule{0.15mm}{0em}{0em}
    % \specialrule{0.15mm}{.1em}{0em}
% \hline
    % \specialrule{0.35mm}{0em}{0em}
\begin{table}
\centering
\caption{Fix patterns of \BugNames}
\label{tab:fix_patterns}
 \resizebox{\linewidth}{!}
 {
\begin{tabular}{c|c|c|c} 
% \hline
\specialrule{0.35mm}{0em}{0em}
\textbf{Interaction Type}                                                            & \textbf{Root Cause}                                       & \textbf{Fix Pattern}                                                                                       & \textbf{\#Bugs}  \\ 
% \hline
\specialrule{0.15mm}{0em}{0em}
\specialrule{0.15mm}{.1em}{0em}
\multirow{15}{*}{\begin{tabular}[c]{@{}c@{}}[A] Intra-system \\Interaction\end{tabular}} & \multirow{2}{*}{{[}A.1.1.1] Namespace mishandling (7)}    & Add namespace passing                                                                                      & 3                \\ 
\cline{3-4}
                                                                                     &                                                           & Fix namespace prefix error                                                                                 & 2                \\ 
\cline{2-4}
                                                                                     & {[}A.1.1.2] Naming mismatch (8)                           & Fix corresponding names                                                                                    & 8                \\ 
\cline{2-4}
                                                                                     & \multirow{2}{*}{{[}A.1.1.3] Incorrect establishment (8)}  & Add/Remove publishers                                                                                      & 3                \\ 
\cline{3-4}
                                                                                     &                                                           & Fix node configurations                                                                                    & 3                \\ 
\cline{2-4}
                                                                                     & {[}A.1.2] Parameter mishandling (7)                       & Add use of declared parameter                                                                            & 3                \\ 
\cline{2-4}
                                                                                     & {[}A.2.1.1] Action goal state mishandling (4)             & Add state check                                                                                            & 2                \\ 
\cline{2-4}
                                                                                     & \multirow{2}{*}{{[}A.2.2] Improper synchronization (12)}  & Add synchronization mechanism                                                                              & 7                \\ 
\cline{3-4}
                                                                                     &                                                           & \begin{tabular}[c]{@{}c@{}}Adjust the timing of\\service/publisher creation\end{tabular}                   & 3                \\ 
\cline{2-4}
                                                                                     & {[}A.2.3.1] Improper node registration (7)                & \begin{tabular}[c]{@{}c@{}}Employ a separate thread for \\node registration and executor spin\end{tabular} & 4                \\ 
\cline{2-4}
                                                                                     & {[}A.2.3.2] Improper callback scheduling settings (4)     & \begin{tabular}[c]{@{}c@{}}Use multi-threaded executor \\and different callback groups\end{tabular}        & 2                \\ 
\cline{2-4}
                                                                                     & \multirow{2}{*}{{[}A.3.1] Definition violation (13)}      & Add assignment for missing fields                                                                          & 5                \\ 
\cline{3-4}
                                                                                     &                                                           & Fix format                                                                                                 & 3                \\ 
\cline{2-4}
                                                                                     & {[}A.3.2] Wrong assumption (4)                            & Fix the usage of received information                                                                      & 4                \\ 
\cline{2-4}
                                                                                     & \multicolumn{2}{c|}{\textbf{Subtotal (83)}}                                                                                                                            & \textbf{52}      \\ 
\hline
\multirow{5}{*}{\begin{tabular}[c]{@{}c@{}}[B] Hardware\\Interaction\end{tabular}}       & \multirow{2}{*}{{[}B.1] Improper resource management (9)} & Fix hardware lifecycle configurations                                                                      & 3                \\ 
\cline{3-4}
                                                                                     &                                                           & Add hardware lifecycle check                                                                               & 3                \\ 
\cline{2-4}
                                                                                     & {[}B.2] Incorrect hardware interfaces (5)                 & Fix hardware interface configurations                                                                      & 3                \\ 
\cline{2-4}
                                                                                     & {[}B.3] Hardware heterogeneity (3)                        & Adjust software                                                                                            & 2                \\ 
\cline{2-4}
                                                                                     & \multicolumn{2}{c|}{\textbf{Subtotal (18)}}                                                                                                                            & \textbf{11}      \\ 
\hline
\multirow{5}{*}{\begin{tabular}[c]{@{}c@{}}[C] Environmental\\Interaction\end{tabular}}  & {[}C.1] Improper (re)planning (4)                         & Add map-related check                                                                                      & 2                \\ 
\cline{2-4}
                                                                                     & \multirow{2}{*}{{[}C.2] Improper collision detection (9)} & Fix collision geometry                                                                                     & 3                \\ 
\cline{3-4}
                                                                                     &                                                           & Fix lookahead distance                                                                                     & 2                \\ 
\cline{2-4}
                                                                                     & {[}C.4] Unexpected environmental conditions (3)           & Add boundary condition handling                                                                            & 3                \\ 
\cline{2-4}
                                                                                     & \multicolumn{2}{c|}{\textbf{Subtotal (20)}}                                                                                                                            & \textbf{10}      \\ 
\hline
\multicolumn{3}{c|}{\textbf{Total (121)}}                                                                                                                                                                                                                     & \textbf{73}      \\
% \hline
\specialrule{0.35mm}{0em}{0em}
\end{tabular}
}
\end{table}

% Developers formulate specific fixes for \BugNames~based on their root causes. Similarly, we extract the fix patterns based on the root causes, and the results are presented in Table \ref{tab:fix_patterns}. 
We extract the fix patterns based on the root causes of \BugNames~which is similar to how developers formulate specific fixes.
The numbers in parentheses after each root cause, subtotal and total, represent the total number of \BugNames~in that category. The ``\#Bugs'' column indicates the number of \BugNames that are fixed using the corresponding fix pattern. \BugNames~not included in Table \ref{tab:fix_patterns} are fixed case-by-case without common patterns. 
For example, among the 12 \BugNames caused by \textit{[A.2.2] improper synchronization}, 7 are fixed by \textit{adding synchronization mechanism} and 3 are resolved by \textit{adjusting the timing of service/publisher creation}. The remaining 2 \BugNames are fixed case-by-case so they are not included in the table.

\textbf{Overview of fix patterns.} We summarize 22 fix patterns, with 76.19\% (16/21) of the atomic root causes having at least one fix pattern and 60.33\% (73/121) of the \BugNames~fixed with one of the patterns. 
Among the 16 root causes with fix patterns, 6 have two or more extracted fix patterns, while the remaining 10 root causes have only one fix pattern. For example, \BugNames caused by \textit{[A.1.1.1] namespace mishandling} can be fixed either by \textit{adding namespace passing} (i.e., enabling the transmission of the namespace from the launch file or command line to the ROS node) or by \textit{fixing namespace prefix error}. For another category \textit{[A.3.2] wrong assumption}, all associated \BugNames are resolved by \textit{fixing the usage of received information}.

Developers can refer to the fix patterns presented in Table \ref{tab:fix_patterns} to address similar \BugNames and researchers may leverage these experiences to develop automated \BugName resolution tools in the future.

\textbf{Case-by-case fixes.} The remaining 39.67\% (48/121) \BugNames~not included in Table \ref{tab:fix_patterns} are addressed through case-by-case fixes. 23.81\% (5/21) root causes do not have a common fix pattern, i.e., \textit{[A.2.1.2] custom state mishandling (3 bugs)}, \textit{[C.3] improper post-processing of environmental information (3 bugs)}, and \textit{others} belong to three types of interactions respectively (6 bugs for \softwareinteraction, 1 bug for \hardwareinteraction, and 1 bug for \environmentinteraction). Developers have conducted a separate study and analysis for each of these \BugNames~to determine the corresponding fix based on the specific situation with their expertise.

\begin{tcolorbox}[colframe=black!75!white, colback=white!90!gray, boxrule=0.2 mm, arc=1mm, boxsep=0.5mm, width=\linewidth]
\stepcounter{finding}
\textbf{Finding \thefinding}: \textit{22 fix patterns are summarized from 76.19\% (16/21) of the root causes, covering 60.33\% (73/121) of the \BugNames.} 
\end{tcolorbox}

% For instance, the 4 bugs under "Incorrect parameter handling" are fixed differently, i.e., "Add usage of declared parameter"(\href{https://github.com/ros-navigation/navigation2/issues/2785}{Navigation2\#2785}), "Fix usage of parameter"(\href{https://github.com/ros-navigation/navigation2/issues/2686}{Navigation2\#2686}), "Add parameter declaration"(\href{https://github.com/ros-navigation/navigation2/issues/1154}{Navigation2\#1154}), and "Fix parameter update logic"(\href{https://github.com/aerostack2/aerostack2/issues/274}{Aerostack2\#274}).
% For the \BugNames~classified as "Others", the fixes are as diverse as their root causes which makes it difficult to extract common patterns.

\subsection{Bug Impacts}
\label{sec:bug_impact}
\begin{figure}
    \centering
    \includegraphics[width=\linewidth]{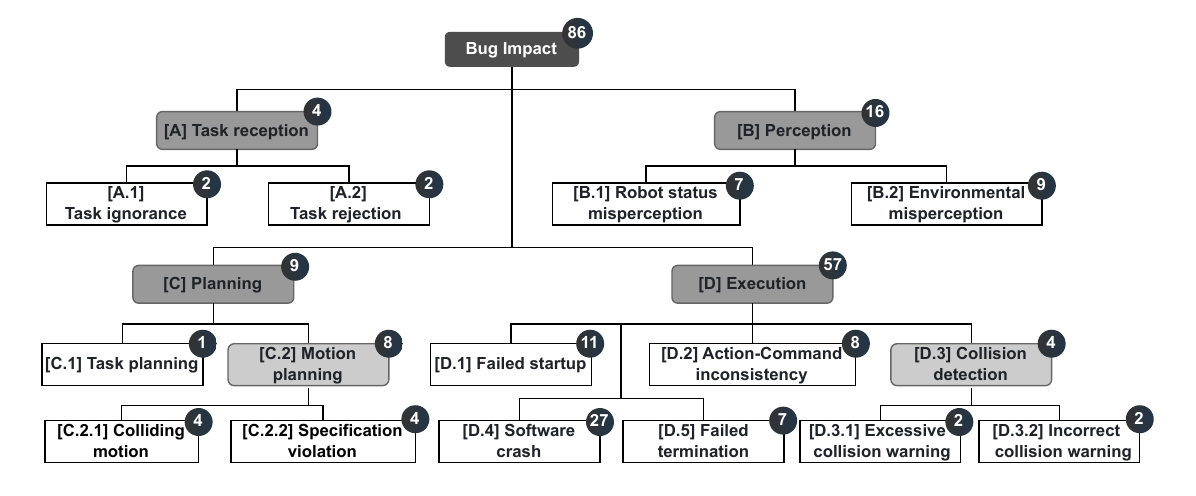}
    \caption{The taxonomy of bug impacts for \BugNames.}
    \label{fig:bug_impact}
\end{figure}

% 这里是不存在double count的，每个bug只被计数一次
% 这里讲例子的时候需要引用前面的bug，提及前面的root cause和fix pattern
We conduct an in-depth study of the bug impacts caused by \BugNames~to better understand how they affect the robotic systems.
For each \BugName, we carefully review its bug report, related discussions, and the source code of the bug to identify the explicit bug impact. The \BugNames~whose impacts cannot be clearly determined due to insufficient information are excluded. Finally, we identify the impacts of \ImpactNumber~\BugNames, accounting for 71.10\% (86/121) of the studied \BugNames. The taxonomy of bug impacts is shown in Figure \ref{fig:bug_impact}, where the meanings of various shapes are similar to those in Figure \ref{fig:root_cause} (see Section \ref{sec:root_cause}).

Our study suggests that \BugNames could impact all core functionalities of the robotic system, including \textit{task reception}, \textit{perception}, \textit{planning}, and \textit{execution}, and thus compromise the robustness and reliability of the robotic system.
Details are introduced as follows.

\textbf{Task reception [A] (4 bugs).} The robot decides whether to accept or reject a task during the \textit{task reception} stage, typically handled by a server node. Four \BugNames cause failures in task reception: two result in \ul{\textit{task ignorance}}, where the tasks sent by users are ignored without any notification, and the remaining two \BugNames lead to \ul{\textit{task rejection}}, where tasks are consistently rejected due to an incorrect transition to the goal state.

\textbf{Perception [B] (16 bugs).} The robotic system collects and analyzes sensor data to comprehend the current status of itself and its surroundings in the \textit{perception} stage. Sixteen \BugNames cause the robot to gain an inaccurate perception. Among these, seven lead to \ul{\textit{robot status misperception}}, which means the robot's awareness of its own state, such as the perceived pose or the bending angles of joints, deviates from reality. The remaining nine \BugNames cause \ul{\textit{environmental misperception}}, wherein the robot's perception of its surroundings, such as the ground depth or map boundaries, does not align with the actual environment.

\textbf{Planning [C] (9 bugs).} The robotic system employs \textit{task planning} \cite{DBLP:journals/ras/GalindoFGS-taskplanning} to sequence high-level actions and \textit{motion planning} \cite{DBLP:journals/isrob/KimKCPOP24-motionplanning} to calculate precise execution paths. Nine \BugNames affect the \textit{planning} stage of the robotic system. Among these, one bug results in wrong behavioral decision-making in \ul{\textit{task planning}} (see \href{https://github.com/ros-navigation/navigation2/issues/1307}{Navigation2\#1307} in Section \ref{subsec:root_cause-environmental_iBugs} [C.1]).

The remaining 8 \BugNames affecting \ul{\textit{motion planning}} cause the robot to generate incorrect execution paths for certain actions. Among them, four \BugNames classified under \ul{\textit{colliding motion}} category produce paths that pass directly through obstacles, leading to collisions. The other four \BugNames, categorized as \ul{\textit{specification violation}}, cause the robot to generate paths that violate the specifications, i.e., the kinematic limitations of the robot and the boundaries of the map.

\textbf{Execution [D] (57 bugs).} 
The robotic system executes the user's tasks in the \textit{execution} stage. 
% The robotic system executes the planned actions and paths in the \textit{execution} stage. 
\BugNames~in this category can affect the entire \textit{execution} stage of the robot, including startup, running, and termination. This category includes  66.28\% (57/86) of the \BugNames with known impacts, making it the most prevalent of the 4 stages.

Eleven \BugNames classified under \ul{\textit{failed startup}} prevent the robot from initiating task execution. Eight \BugNames classified as \ul{\textit{action-command inconsistency}} cause the actual action executed by the robot to be inconsistent with the command sent by the user. For example, the robot in \href{https://github.com/aerostack2/aerostack2/issues/655}{Aerostack2\#655} stops at the first waypoint instead of executing the entire path specified by the user command.

Four \BugNames affects the \ul{\textit{collision detection}} of the robotic system, which detects potential undesirable collisions with the environment. Among these, two \BugNames introduce \ul{\textit{excessive collision warnings}}, meaning that a collision warning is triggered when there is still a sufficient distance between the obstacle and the robot. In such cases, no actual collision will occur if the robot continues to follow its planned path. The other two \BugNames lead to \ul{\textit{incorrect collision warning}}, in which the collision warning is incorrect due to angular or positional deviations between the reference state of the collision detection module and the actual state of the robot.

Twenty-seven \BugNames lead to \ul{\textit{software crashes}} in the robotic system. A crash typically originates from a single node and may propagate to other dependent nodes, potentially leading to severe problems. This category accounts for 31.40\% (27/86) of the \BugNames~ studied in this section and become the most prevalent atomic category among the bug impacts.

Seven \BugNames classified under \ul{\textit{failed termination}} can prevent the robot from properly terminating the executing action. Such a termination could be initiated either by the system itself or by the user. 
For example, the robot in \href{https://github.com/ros-navigation/navigation2/issues/3361}{Navigation2\#3361} becomes stuck at the last velocity command due to an unhandled exception, subsequently entering an unsafe state of persistent motion.

\begin{tcolorbox}[colframe=black!75!white, colback=white!90!gray, boxrule=0.2 mm, arc=1mm, boxsep=0.5mm, width=\linewidth]
\refstepcounter{finding}
\textbf{Finding \thefinding}: \textit{71.10\% (86/121) of the \BugNames~are identified to have an impact on the general process by which a robot receives and executes a task.
Among these, execution constitutes the largest proportion (66.28\%$\approx$57/86).
% i.e., task reception (4.65\%$\approx$4/86), perception (18.60\%$\approx$16/86), planning (10.47\%$\approx$9/86), and execution(66.28\%$\approx$57/86). 
% Their respective proportions are , 18.60\% (16/86), 10.47\% (9/86), and 66.28\% (57/86).
} 
\label{finding:impact}
\end{tcolorbox}

\section{Discussion}
\label{sec:discussion}

\BugNames~can have diverse impacts on the robotic system (Finding \ref{finding:impact}). Ensuring the correctness of interactions is important for the reliability of the robotic system. In this section, we discuss potential measures for developers to avoid \BugNames, implications to existing methods and opportunities for new research in combating \BugNames.

\vspace{-2pt}
\subsection{\BugNames Avoidance}
% Robotic developers should pay more attention to interactions in robotic systems, as they are both prevalent and complex.
Robotic developers should pay more attention to the interactions in the robotic system, as they are both prevalent and complex.

\textbf{Verify the interaction topology.} 
The interaction topology is a common source of \BugNames (Finding \ref{finding:root_cause-topology_error}). Developers, upon completing relevant modifications, are suggested to utilize tools like the rqt plugin \cite{Rqt} to verify whether the establishment of nodes and interfaces, the interconnections among them, and their namespaces align with the expectations.

\textbf{Ensure proper use of executors.} The complex setup and usage of ROS executors affects a lot of \BugNames (Finding \ref{finding:root_cause-executor_misuse}). For nodes requiring callback execution, developers should ensure that each node is registered with the executor only once and that the executor is started to spin. If a node needs to execute a callback function for continuously processing incoming messages, it is advisable to assign a dedicated thread for its executor rather than using the executor each time the messages are needed, thereby preventing potential redundancies or omissions. 
Moreover, when multiple callback functions need to be executed concurrently within the same node, either explicitly or implicitly (e.g., when a callback function performs a blocking wait for a service or action response, with an implicit done-callback), developers should assign each callback to an appropriate callback group and use a multi-threaded executor. Failure to do so may result in deadlocks.

\textbf{Examine the use of ROS interaction interfaces.} 
The customizable nature of data structures within these interfaces contributes to many \BugNames (Finding \ref{finding:root_cause-interface_misuse}). 
% The customizable nature of data structures within these interfaces introduces a risk of bugs, contributing to 14.05\% (17/121) of the total \BugNames (Finding \ref{finding:root_cause-intra_system}). 
Developers should ensure that data structures transmitted through these interfaces (including topics, services, and actions) conform to their definition in configuration files. Specifically, all fields must be correctly assigned and formatted in the sending node, while the receiving node should process the data in accordance with the predefined field types.

\textbf{Pay attention to hardware resource management.} \textit{Improper resource management} accounts for the most hardware \BugNames (Finding \ref{finding:root_cause-hardware}). Developers should ensure that the hardware resources are correctly assigned to their corresponding controllers. Before accessing a hardware component, its lifecycle state should be checked to confirm availability. In case where a hardware component returns an error or throws an exception,  indicating its unavailability, its lifecycle state must be updated accordingly.

\textbf{Ensure effective collision detection.} \textit{Improper collision detection} is the most common root cause among the environmental \BugNames (Finding \ref{finding:root_cause-environment}). Developers should ensure that the collision geometry is properly configured to accurately represent the actual shape of both the robot and its environment. Furthermore, the look-ahead direction should be aligned with the robot’s planned motion to minimize angular deviations. The computed look-ahead point should not exceed the map boundaries and the robot’s planned path to avoid memory faults and unwanted warnings, respectively.

\vspace{-3pt}
\subsection{\BugNames Detection}
Resolving \BugNames~is of great significance for the reliability of the robotic system. Our study summarizes three types of interactions and 21 atomic categories of root causes for \BugNames, shedding new light and guidance on \BugName~detection.

% patterns 可以基于Root cause来获取:消息的几种bug；状态的几种bug，状态转换前缺少检查，状态转换不符合状态机，进行某些操作时缺少状态转换；ROSDiscover可以检测topic name mismatch，这覆盖了其中的一种，以及namespace中会产生Mismatch的一部分；对于剩余的部分，比如代码中遗漏了对namespace的支持以及等待了错误的命名空间中的服务，则需要对额外的工作； 这里每次描述完Pattern，就说下对应的检测工具大概要怎么做

% 
% \textbf{Pattern-based bug detection.} 

\textbf{\Softwareinteraction~bug detection.} The explicit patterns of intra-system \BugNames can be extracted from root causes, providing a basis for developing static analysis tools for bug detection. 
For example, the patterns of \textit{definition violation} include omitting assignments and constructing incorrect formats for certain fields within the data structures of ROS interaction interfaces. Static analysis tools can be designed to detect whether these assignments conform to the definitions declared in the configuration files. 
In addition, type-checking tools can be adapted to detect whether received data fields are used in accordance with their defined types. 
ROSDiscover \cite{DBLP:conf/icsa/TimperleyDSGG22-ROSDiscover} can detect dangling subscribers/publishers, i.e., a subscriber without a publisher and vice versa, resulted from \textit{interaction topology error}. Further improvements to ROSDiscover are still needed to detect \BugNames from this category that do not cause dangling. 

% 现有的动态工作，由于它们将用于触发ROS系统内交互的接口，比如topic message，service request的全部或部分纳入输入空间，因此有可能检测出交互bug。但用的是通用预言机，只能检测一些并发、内存和崩溃相关的bug；对于交互的正确性，比如任务请求一直被拒绝由于状态转换错误，还需要对Oracle进行额外的改变
Existing fuzzing tools designed for ROS \cite{ROS2Fuzz,DBLP:conf/issta/ShenLXSWGS024-FuzzingCallback,DBLP:conf/icra/XieBZ022-ROZZ-Fuzz,DBLP:conf/asplos/BaiS024-ROFER-Fuzz} incorporate some or all data structures within ROS interaction interfaces, such as topic messages and service requests, in their input space, making them possible to detect some of the intra-system \BugNames. These tools employ general oracles that capture bugs related to concurrency, memory, and crashes. Further enhancements to their oracles are still required to detect \BugNames that affect the correctness of \softwareinteraction s without general features, such as task requests being consistently rejected due to \textit{state mishandling}.

%静态分析，判断和文档中所描述的向硬件发送的指令或者操作是否相符，与接口名称是否相符。动态分析，简单地执行就可检测出使接口不可用的bug
\textbf{\Hardwareinteraction~bug detection.} Static tools can be designed to detect whether the hardware interface implementations align with their documented specifications and naming conventions. For instance, such tools can identify the \BugName in which a hardware interface mistakenly sends velocity commands to the wheels while its naming and documentation suggest that it is a position command interface (\href{https://github.com/ros-controls/ros2_controllers/issues/1167}{ROS2\_controllers\#1167}).
% 以硬件的状态转换（覆盖）为指引生成测试用例，比如开启、关闭硬件，甚至注入硬件错误，测试硬件资源管理是否可靠
In terms of dynamic analysis, test case generations can be guided by the state transitions of the hardware within its lifecycle. For instance, test cases can be designed to activate and deactivate hardware devices, as well as inject hardware faults, to maximize coverage of hardware states and assess the reliability of hardware resource management.

\textbf{\Environmentinteraction~bug detection.} Environmental \BugNames are often triggered by complex and dynamic physical environments, making them difficult to detect through static analysis except in cases of fixed bug patterns, such as missing boundary checks during map access.
Extensive testing in diverse conditions, whether in simulators or real environments, is suggested. RoboFuzz \cite{DBLP:conf/sigsoft/KimK22-RoboFuzz} performs fuzz testing on ROS systems with an extendable oracle to detect correctness bugs during operation. By manually extending RoboFuzz's oracle to define the correct behavior of the robot, it can detect some predictable environmental \BugNames such as when a robot fails to stop upon encountering an obstacle. 
% The challenge lies in the significant effort required to manually construct oracles and the inherent difficulty of ensuring comprehensive coverage of all possible scenarios.
The challenge stems from the significant manual effort involved in oracle construction, combined with the difficulty of achieving thorough coverage across various situations.
% The challenge lies in that oracle construction requires significant manual effort, and it is difficult to cover all possible situations. 
% The challenge lies in that constructing oracles requires significant effort and it is difficult to cover all possible situations. 
Automatically extracting oracles from robot specifications is a promising direction for detecting environmental \BugNames, warranting further research in the future.
% Developing automated oracles from robot specifications is a promising direction for detecting environmental \BugNames, warranting further research in the future.

\vspace{-3pt}
\section{Related Work}
\label{sec:related_work}

% In this section, we review existing studies, which are organized along two main perspectives: empirical studies on robotic systems and bug detection approaches for robotic systems.

\textbf{Empirical studies on robotic systems.} Researchers have conducted a series of empirical studies to deepen the understanding of robotic systems. Kolak et al. \cite{DBLP:conf/icsm/KolakAGHT20-village} conducted an empirical study on the ROS ecosystem. They analyzed the participants, the dependencies between software packages, and the overall structure of the ROS ecosystem.
Fischer-Nielsen et al. \cite{DBLP:conf/icse/Fischer-Nielsen20-ROSDependency} studied dependency bugs in ROS, i.e., software faults caused by incorrect references to external software assets, and presented the associated definition and taxonomy. 
Malavolta et al. \cite{DBLP:conf/icse/MalavoltaLSLG20-architect} studied the architectural practices of ROS-based systems. 
Cottrell et al. \cite{DBLP:conf/compsac/CottrellBSR21-ROSvul} analyzed 176 vulnerabilities related to robots and classified them into nine major categories. Song et al. \cite{DBLP:conf/icsm/SongLDLCP23-FaultDiagnosis} investigated the effectiveness of using observability data, i.e. logs, traces, and trajectories, for fault diagnosis in ROS-based systems. 
Timperley et al. \cite{DBLP:journals/ese/TimperleyHSDW24-ROBUST221bugs} studied general bugs across seven ROS-based systems. They categorized bug faults into seven coarse-grained categories and further analyzed bug failures and fixes.
Canelas et al. \cite{DBLP:conf/issta/CanelasSFT24-misconfiguration} proposed a taxonomy for misconfigurations in ROS and theoretically analyzed whether existing methods could detect the bugs in their dataset. 
Unlike previous empirical studies, our research provides a comprehensive analysis of \BugNames~ arising from interactions both among components within the robotic system and between the system and the environment.
% that arise from the interactions between different components in the robotic system as well as the interactions between the robotic system and the environment. 

\textbf{Detecting bugs in robotic systems.} A variety of bug detection approaches have been proposed to improve the reliability of robotic systems.  
% correspondence：对应关系
Kate et al. \cite{DBLP:conf/sigsoft/KateCCZE21-physframe} proposed PHYSFRAME, which employs static analysis by modeling the correspondence between variables and physical reference frames as types, and defining a series of type-checking rules to detect bugs related to inconsistencies in physical reference frames. 
ROSDiscover \cite{DBLP:conf/icsa/TimperleyDSGG22-ROSDiscover} is a static method for run-time architecture misconfiguration detection in robotic systems. It can be employed to detect certain intra-system \BugNames that result in dangling subscribers and publishers. 
ROSInfer \cite{DBLP:conf/icse/DurschmidTGG24-ROSInfer} goes a step further by modeling the internal behavior of nodes as state machines, which enables it to detect architecture composition bugs that affect the internal behavior of nodes.

In addition to the above static methods, fuzzing approaches \cite{ROS2Fuzz,DBLP:conf/icra/XieBZ022-ROZZ-Fuzz,DBLP:conf/asplos/BaiS024-ROFER-Fuzz,DBLP:conf/issta/ShenLXSWGS024-FuzzingCallback,DBLP:conf/sigsoft/KimK22-RoboFuzz} have been proposed for robotic systems. Their difference includes input space, mutation strategies, feedback metrics, and test oracles. Among these, \cite{ROS2Fuzz,DBLP:conf/issta/ShenLXSWGS024-FuzzingCallback,DBLP:conf/icra/XieBZ022-ROZZ-Fuzz,DBLP:conf/asplos/BaiS024-ROFER-Fuzz} employ general oracles which primarily focus on bugs related to concurrency, memory, and crashes. 
% Some of their inputs related to intra-system interactions like topic messages and service requests may be able to trigger limited \BugNames. 
Some of their inputs related to intra-system interactions, such as topic messages and service requests, may be capable of triggering certain \BugNames; however, any \BugNames that cannot be captured by their oracles will remain undetected.
The oracle of RoboFuzz \cite{DBLP:conf/sigsoft/KimK22-RoboFuzz} can be extended through extensive manual effort to detect some predictable environmental \BugNames. However, manual efforts are limited in covering numerous possible scenarios.
Despite these methods, there is no dedicated one for detecting hardware \BugNames.

\vspace{-2pt}
\section{Conclusion}
\label{sec:conclusion}

% In modern society, the advent of robotics has revolutionized various industries, from manufacturing to healthcare, and beyond.
The reliability of robotic systems is a critical factor that affects not only their performance, but also their trustworthiness in the eyes of the users.
In this paper, we conduct the first empirical study to systematically investigate the interaction bugs (\BugNames) in robotic systems, which is a common and important failure mode.
Within the ecosystem of Robot Operating System (ROS), we identify \FullBugNumber \BugNames and carefully analyze \BugNumber of them to understand their root causes, fix strategies, and bug impacts.
Our study shows that the modular architecture of robotic systems complicates the management of intra-system communications in terms of configuration, synchronization, etc.
Moreover, developers need to carefully manage device resources and account for their heterogeneity when integrating software and hardware components, as well as effectively handle the dynamic nature of environments during external interactions.
% Moreover, developers need to carefully handle device heterogeneity when integrating software and hardware components, as well as the dynamic nature of environments during external interactions.
These bugs can impact all core functionalities of robots, including task reception, perception, planning, and execution.
Based on our findings, we discuss practical directions for \BugNames avoidance and detection.
We believe our work can shed light on advancing the reliability of robotic systems.
% We believe our work can shed light on advancing the performance and safety of robotic systems.

%%marefs
% \begin{acks}
% \end{acks}

\bibliographystyle{ACM-Reference-Format}
\balance
\bibliography{refs}


\begin{thebibliography}{45}

%%% the Web address.

\ifx \showCODEN    \undefined \def \showCODEN     #1{\unskip}     \fi
\ifx \showDOI      \undefined \def \showDOI       #1{#1}\fi
\ifx \showISBNx    \undefined \def \showISBNx     #1{\unskip}     \fi
\ifx \showISBNxiii \undefined \def \showISBNxiii  #1{\unskip}     \fi
\ifx \showISSN     \undefined \def \showISSN      #1{\unskip}     \fi
\ifx \showLCCN     \undefined \def \showLCCN      #1{\unskip}     \fi
\ifx \shownote     \undefined \def \shownote      #1{#1}          \fi
\ifx \showarticletitle \undefined \def \showarticletitle #1{#1}   \fi
\ifx \showURL      \undefined \def \showURL       {\relax}        \fi
% invisible to TeX
\providecommand\bibfield[2]{#2}
\providecommand\bibinfo[2]{#2}
\providecommand\natexlab[1]{#1}
\providecommand\showeprint[2][]{arXiv:#2}

\bibitem[our(2025)]%
        {our_dataset}
 \bibinfo{year}{2025}\natexlab{}.
\newblock \bibinfo{booktitle}{\emph{Dataset of ``Understanding Interaction Bugs in Robotic Systems''}}.
\newblock
\urldef\tempurl%
\url{https://anonymous.4open.science/r/Understanding-Interaction-Bugs-in-Robotic-Systems-FC8E/README.md}
\showURL{%
Retrieved March 13, 2025 from \tempurl}


\bibitem[AndreasAZiegler(2019)]%
        {NavIssue-continues-moving-when-cancel}
\bibfield{author}{\bibinfo{person}{AndreasAZiegler}.} \bibinfo{year}{2019}\natexlab{}.
\newblock \bibinfo{booktitle}{\emph{Navigation2 Issue 911: Robot continues moving when pressing ``cancel'' button during navigation.}}
\newblock
\urldef\tempurl%
\url{https://github.com/ros-navigation/navigation2/issues/911}
\showURL{%
Retrieved October 31, 2024 from \tempurl}


\bibitem[Anumandla and Tejani(2023)]%
        {anumandla2023robotic-advantage}
\bibfield{author}{\bibinfo{person}{Sunil Kumar~Reddy Anumandla} {and} \bibinfo{person}{JG Tejani}.} \bibinfo{year}{2023}\natexlab{}.
\newblock \showarticletitle{Robotic Automation in Rubber Processing: Improving Safety and Productivity}.
\newblock \bibinfo{journal}{\emph{Asian Journal of Applied Science and Engineering}} \bibinfo{volume}{12}, \bibinfo{number}{1} (\bibinfo{year}{2023}), \bibinfo{pages}{7--15}.
\newblock


\bibitem[Awad et~al\mbox{.}(2016)]%
        {DBLP:conf/etfa/AwadHRB16-workbench}
\bibfield{author}{\bibinfo{person}{Ramez Awad}, \bibinfo{person}{Georg Heppner}, \bibinfo{person}{Arne Roennau}, {and} \bibinfo{person}{Mirko Bordignon}.} \bibinfo{year}{2016}\natexlab{}.
\newblock \showarticletitle{{ROS} engineering workbench based on semantically enriched app models for improved reusability}. In \bibinfo{booktitle}{\emph{21st {IEEE} International Conference on Emerging Technologies and Factory Automation, {ETFA} 2016, 2016}}. \bibinfo{publisher}{{IEEE}}, \bibinfo{pages}{1--9}.
\newblock
\urldef\tempurl%
\url{https://doi.org/10.1109/ETFA.2016.7733581}
\showDOI{\tempurl}


\bibitem[Bai et~al\mbox{.}(2024)]%
        {DBLP:conf/asplos/BaiS024-ROFER-Fuzz}
\bibfield{author}{\bibinfo{person}{Jia{-}Ju Bai}, \bibinfo{person}{Haoxuan Song}, {and} \bibinfo{person}{Shimin Hu}.} \bibinfo{year}{2024}\natexlab{}.
\newblock \showarticletitle{Multi-Dimensional and Message-Guided Fuzzing for Robotic Programs in Robot Operating System}. In \bibinfo{booktitle}{\emph{Proceedings of the 29th {ACM} International Conference on Architectural Support for Programming Languages and Operating Systems, Volume 2, {ASPLOS} 2024}}. \bibinfo{publisher}{{ACM}}, \bibinfo{pages}{763--778}.
\newblock
\urldef\tempurl%
\url{https://doi.org/10.1145/3620665.3640425}
\showDOI{\tempurl}


\bibitem[Biggs et~al\mbox{.}(2020)]%
        {ROS2Design-Goal-states}
\bibfield{author}{\bibinfo{person}{Geoffrey Biggs}, \bibinfo{person}{Jacob Perron}, {and} \bibinfo{person}{Shane Loretz}.} \bibinfo{year}{2020}\natexlab{}.
\newblock \bibinfo{booktitle}{\emph{ROS 2 Design: Actions}}.
\newblock
\urldef\tempurl%
\url{https://design.ros2.org/articles/actions.html\#goal-states}
\showURL{%
Retrieved October 31, 2024 from \tempurl}


\bibitem[Bouraine et~al\mbox{.}(2022)]%
        {DBLP:journals/evi/Robucar}
\bibfield{author}{\bibinfo{person}{Sara Bouraine}, \bibinfo{person}{Abdelhak Bougouffa}, {and} \bibinfo{person}{Ouahiba Azouaoui}.} \bibinfo{year}{2022}\natexlab{}.
\newblock \showarticletitle{Particle swarm optimization for solving a scan-matching problem based on the normal distributions transform}.
\newblock \bibinfo{journal}{\emph{Evol. Intell.}} \bibinfo{volume}{15}, \bibinfo{number}{1} (\bibinfo{year}{2022}), \bibinfo{pages}{683--694}.
\newblock
\urldef\tempurl%
\url{https://doi.org/10.1007/S12065-020-00545-Y}
\showDOI{\tempurl}


\bibitem[Brooks et~al\mbox{.}(2005a)]%
        {DBLP:conf/icra/BrooksKMWO05-Orca}
\bibfield{author}{\bibinfo{person}{Alex Brooks}, \bibinfo{person}{Tobias Kaupp}, \bibinfo{person}{Alexei Makarenko}, \bibinfo{person}{Stefan~B. Williams}, {and} \bibinfo{person}{Anders Oreb{\"{a}}ck}.} \bibinfo{year}{2005}\natexlab{a}.
\newblock \showarticletitle{Orca: {A} Component Model and Repository}. In \bibinfo{booktitle}{\emph{Software Engineering for Experimental Robotics, Workshop on Principles and Practice of Software Development in Robotics, PPSDR@ICRA 2005}} \emph{(\bibinfo{series}{Springer Tracts in Advanced Robotics}, Vol.~\bibinfo{volume}{30})}. \bibinfo{pages}{231--251}.
\newblock
\urldef\tempurl%
\url{https://doi.org/10.1007/978-3-540-68951-5\_13}
\showDOI{\tempurl}


\bibitem[Brooks et~al\mbox{.}(2005b)]%
        {DBLP:conf/iros/BrooksKMWO05-component-based-robotics}
\bibfield{author}{\bibinfo{person}{Alex Brooks}, \bibinfo{person}{Tobias Kaupp}, \bibinfo{person}{Alexei Makarenko}, \bibinfo{person}{Stefan~B. Williams}, {and} \bibinfo{person}{Anders Oreb{\"{a}}ck}.} \bibinfo{year}{2005}\natexlab{b}.
\newblock \showarticletitle{Towards component-based robotics}. In \bibinfo{booktitle}{\emph{2005 {IEEE/RSJ} International Conference on Intelligent Robots and Systems, 2005}}. \bibinfo{publisher}{{IEEE}}, \bibinfo{pages}{163--168}.
\newblock
\urldef\tempurl%
\url{https://doi.org/10.1109/IROS.2005.1545523}
\showDOI{\tempurl}


\bibitem[Brugali and Scandurra(2009)]%
        {DBLP:journals/ram/BrugaliS09-component-based-robotics}
\bibfield{author}{\bibinfo{person}{Davide Brugali} {and} \bibinfo{person}{Patrizia Scandurra}.} \bibinfo{year}{2009}\natexlab{}.
\newblock \showarticletitle{Component-based robotic engineering (Part {I)} [Tutorial]}.
\newblock \bibinfo{journal}{\emph{{IEEE} Robotics Autom. Mag.}} \bibinfo{volume}{16}, \bibinfo{number}{4} (\bibinfo{year}{2009}), \bibinfo{pages}{84--96}.
\newblock
\urldef\tempurl%
\url{https://doi.org/10.1109/MRA.2009.934837}
\showDOI{\tempurl}


\bibitem[Brugali and Shakhimardanov(2010)]%
        {DBLP:journals/ram/BrugaliS10-component-based-robotics}
\bibfield{author}{\bibinfo{person}{Davide Brugali} {and} \bibinfo{person}{Azamat Shakhimardanov}.} \bibinfo{year}{2010}\natexlab{}.
\newblock \showarticletitle{Component-Based Robotic Engineering (Part {II)}}.
\newblock \bibinfo{journal}{\emph{{IEEE} Robotics Autom. Mag.}} \bibinfo{volume}{17}, \bibinfo{number}{1} (\bibinfo{year}{2010}), \bibinfo{pages}{100--112}.
\newblock
\urldef\tempurl%
\url{https://doi.org/10.1109/MRA.2010.935798}
\showDOI{\tempurl}


\bibitem[Canelas et~al\mbox{.}(2024)]%
        {DBLP:conf/issta/CanelasSFT24-misconfiguration}
\bibfield{author}{\bibinfo{person}{Paulo Canelas}, \bibinfo{person}{Bradley~R. Schmerl}, \bibinfo{person}{Alcides Fonseca}, {and} \bibinfo{person}{Christopher~Steven Timperley}.} \bibinfo{year}{2024}\natexlab{}.
\newblock \showarticletitle{Understanding Misconfigurations in {ROS:} An Empirical Study and Current Approaches}. In \bibinfo{booktitle}{\emph{Proceedings of the 33rd {ACM} {SIGSOFT} International Symposium on Software Testing and Analysis, {ISSTA} 2024, 2024}}. \bibinfo{publisher}{{ACM}}, \bibinfo{pages}{1161--1173}.
\newblock
\urldef\tempurl%
\url{https://doi.org/10.1145/3650212.3680350}
\showDOI{\tempurl}


\bibitem[Cottrell et~al\mbox{.}(2021)]%
        {DBLP:conf/compsac/CottrellBSR21-ROSvul}
\bibfield{author}{\bibinfo{person}{Kaitlyn Cottrell}, \bibinfo{person}{Dibyendu~Brinto Bose}, \bibinfo{person}{Hossain Shahriar}, {and} \bibinfo{person}{Akond Rahman}.} \bibinfo{year}{2021}\natexlab{}.
\newblock \showarticletitle{An Empirical Study of Vulnerabilities in Robotics}. In \bibinfo{booktitle}{\emph{{IEEE} 45th Annual Computers, Software, and Applications Conference, {COMPSAC} 2021, 2021}}. \bibinfo{publisher}{{IEEE}}, \bibinfo{pages}{735--744}.
\newblock
\urldef\tempurl%
\url{https://doi.org/10.1109/COMPSAC51774.2021.00105}
\showDOI{\tempurl}


\bibitem[D{\"{u}}rschmid et~al\mbox{.}(2024)]%
        {DBLP:conf/icse/DurschmidTGG24-ROSInfer}
\bibfield{author}{\bibinfo{person}{Tobias D{\"{u}}rschmid}, \bibinfo{person}{Christopher~Steven Timperley}, \bibinfo{person}{David Garlan}, {and} \bibinfo{person}{Claire {Le Goues}}.} \bibinfo{year}{2024}\natexlab{}.
\newblock \showarticletitle{ROSInfer: Statically Inferring Behavioral Component Models for ROS-based Robotics Systems}. In \bibinfo{booktitle}{\emph{Proceedings of the 46th {IEEE/ACM} International Conference on Software Engineering, {ICSE} 2024}}. \bibinfo{publisher}{{ACM}}, \bibinfo{pages}{144:1--144:13}.
\newblock
\urldef\tempurl%
\url{https://doi.org/10.1145/3597503.3639206}
\showDOI{\tempurl}


\bibitem[Estefo et~al\mbox{.}(2019)]%
        {DBLP:journals/jss/EstefoSRF19-package-reuse}
\bibfield{author}{\bibinfo{person}{Pablo Estefo}, \bibinfo{person}{Jocelyn Simmonds}, \bibinfo{person}{Romain Robbes}, {and} \bibinfo{person}{Johan Fabry}.} \bibinfo{year}{2019}\natexlab{}.
\newblock \showarticletitle{The Robot Operating System: Package reuse and community dynamics}.
\newblock \bibinfo{journal}{\emph{J. Syst. Softw.}}  \bibinfo{volume}{151} (\bibinfo{year}{2019}), \bibinfo{pages}{226--242}.
\newblock
\urldef\tempurl%
\url{https://doi.org/10.1016/J.JSS.2019.02.024}
\showDOI{\tempurl}


\bibitem[Faheem et~al\mbox{.}(2024)]%
        {faheem2024ai-advantage}
\bibfield{author}{\bibinfo{person}{Muhammad~Ashraf Faheem}, \bibinfo{person}{Nabeel Zafar}, \bibinfo{person}{Parkash Kumar}, \bibinfo{person}{Md~Mehedi~Hassan Melon}, \bibinfo{person}{Nayem~Uddin Prince}, {and} \bibinfo{person}{Mohd~Abdullah Al~Mamun}.} \bibinfo{year}{2024}\natexlab{}.
\newblock \showarticletitle{AI AND ROBOTIC: ABOUT THE TRANSFORMATION OF CONSTRUCTION INDUSTRY AUTOMATION AS WELL AS LABOR PRODUCTIVITY}.
\newblock \bibinfo{journal}{\emph{Remittances Review}} \bibinfo{volume}{9}, \bibinfo{number}{S3 (July 2024)} (\bibinfo{year}{2024}), \bibinfo{pages}{871--888}.
\newblock


\bibitem[Fischer{-}Nielsen et~al\mbox{.}(2020)]%
        {DBLP:conf/icse/Fischer-Nielsen20-ROSDependency}
\bibfield{author}{\bibinfo{person}{Anders Fischer{-}Nielsen}, \bibinfo{person}{Zhoulai Fu}, \bibinfo{person}{Ting Su}, {and} \bibinfo{person}{Andrzej Wasowski}.} \bibinfo{year}{2020}\natexlab{}.
\newblock \showarticletitle{The forgotten case of the dependency bugs: on the example of the robot operating system}. In \bibinfo{booktitle}{\emph{{ICSE-SEIP} 2020: 42nd International Conference on Software Engineering, Software Engineering in Practice, 2020}}. \bibinfo{publisher}{{ACM}}, \bibinfo{pages}{21--30}.
\newblock
\urldef\tempurl%
\url{https://doi.org/10.1145/3377813.3381364}
\showDOI{\tempurl}


\bibitem[Future(2024)]%
        {RoboticMarket}
\bibfield{author}{\bibinfo{person}{Market~Research Future}.} \bibinfo{year}{2024}\natexlab{}.
\newblock \bibinfo{booktitle}{\emph{Robotics Market Research Report}}.
\newblock
\urldef\tempurl%
\url{https://www.marketresearchfuture.com/reports/robotics-market-4732}
\showURL{%
Retrieved October 30, 2024 from \tempurl}


\bibitem[Galindo et~al\mbox{.}(2008)]%
        {DBLP:journals/ras/GalindoFGS-taskplanning}
\bibfield{author}{\bibinfo{person}{Cipriano Galindo}, \bibinfo{person}{Juan{-}Antonio Fern{\'{a}}ndez{-}Madrigal}, \bibinfo{person}{Javier Gonz{\'{a}}lez}, {and} \bibinfo{person}{Alessandro Saffiotti}.} \bibinfo{year}{2008}\natexlab{}.
\newblock \showarticletitle{Robot task planning using semantic maps}.
\newblock \bibinfo{journal}{\emph{Robotics Auton. Syst.}} \bibinfo{volume}{56}, \bibinfo{number}{11} (\bibinfo{year}{2008}), \bibinfo{pages}{955--966}.
\newblock
\urldef\tempurl%
\url{https://doi.org/10.1016/J.ROBOT.2008.08.007}
\showDOI{\tempurl}


\bibitem[Garc{\'{\i}}a et~al\mbox{.}(2020)]%
        {DBLP:conf/sigsoft/00020BBP20-service-robot}
\bibfield{author}{\bibinfo{person}{Sergio Garc{\'{\i}}a}, \bibinfo{person}{Daniel Str{\"{u}}ber}, \bibinfo{person}{Davide Brugali}, \bibinfo{person}{Thorsten Berger}, {and} \bibinfo{person}{Patrizio Pelliccione}.} \bibinfo{year}{2020}\natexlab{}.
\newblock \showarticletitle{Robotics software engineering: a perspective from the service robotics domain}. In \bibinfo{booktitle}{\emph{{ESEC/FSE} '20: 28th {ACM} Joint European Software Engineering Conference and Symposium on the Foundations of Software Engineering, 2020}}. \bibinfo{publisher}{{ACM}}, \bibinfo{pages}{593--604}.
\newblock
\urldef\tempurl%
\url{https://doi.org/10.1145/3368089.3409743}
\showDOI{\tempurl}


\bibitem[Gerkey et~al\mbox{.}(2001)]%
        {DBLP:conf/iros/GerkeyVSHSM01-player}
\bibfield{author}{\bibinfo{person}{Brian~P. Gerkey}, \bibinfo{person}{Richard~T. Vaughan}, \bibinfo{person}{Kasper St{\o}y}, \bibinfo{person}{Andrew Howard}, \bibinfo{person}{Gaurav~S. Sukhatme}, {and} \bibinfo{person}{Maja~J. Mataric}.} \bibinfo{year}{2001}\natexlab{}.
\newblock \showarticletitle{Most valuable player: a robot device server for distributed control}. In \bibinfo{booktitle}{\emph{{IEEE/RSJ} International Conference on Intelligent Robots and Systems, {IROS} 2001: Expanding the Societal Role of Robotics in the the Next Millennium, 2001}}. \bibinfo{publisher}{{IEEE}}, \bibinfo{pages}{1226--1231}.
\newblock
\urldef\tempurl%
\url{https://doi.org/10.1109/IROS.2001.977150}
\showDOI{\tempurl}


\bibitem[Jahn et~al\mbox{.}(2019)]%
        {DBLP:journals/computers/JahnWS19-modular-robotic}
\bibfield{author}{\bibinfo{person}{Uwe Jahn}, \bibinfo{person}{Carsten Wolff}, {and} \bibinfo{person}{Peter Schulz}.} \bibinfo{year}{2019}\natexlab{}.
\newblock \showarticletitle{Concepts of a Modular System Architecture for Distributed Robotic Systems}.
\newblock \bibinfo{journal}{\emph{Comput.}} \bibinfo{volume}{8}, \bibinfo{number}{1} (\bibinfo{year}{2019}), \bibinfo{pages}{25}.
\newblock
\urldef\tempurl%
\url{https://doi.org/10.3390/COMPUTERS8010025}
\showDOI{\tempurl}


\bibitem[Javaid et~al\mbox{.}(2021)]%
        {javaid2021substantial-advantage}
\bibfield{author}{\bibinfo{person}{Mohd Javaid}, \bibinfo{person}{Abid Haleem}, \bibinfo{person}{Ravi~Pratap Singh}, {and} \bibinfo{person}{Rajiv Suman}.} \bibinfo{year}{2021}\natexlab{}.
\newblock \showarticletitle{Substantial capabilities of robotics in enhancing industry 4.0 implementation}.
\newblock \bibinfo{journal}{\emph{Cognitive Robotics}}  \bibinfo{volume}{1} (\bibinfo{year}{2021}), \bibinfo{pages}{58--75}.
\newblock


\bibitem[JnxF and gavanderhoorn(2021)]%
        {ROS2Fuzz}
\bibfield{author}{\bibinfo{person}{JnxF} {and} \bibinfo{person}{gavanderhoorn}.} \bibinfo{year}{2021}\natexlab{}.
\newblock \bibinfo{booktitle}{\emph{ros2\_fuzz}}.
\newblock
\urldef\tempurl%
\url{https://github.com/rosin-project/ros2_fuzz}
\showURL{%
Retrieved October 23, 2024 from \tempurl}


\bibitem[Kate et~al\mbox{.}(2021)]%
        {DBLP:conf/sigsoft/KateCCZE21-physframe}
\bibfield{author}{\bibinfo{person}{Sayali Kate}, \bibinfo{person}{Michael Chinn}, \bibinfo{person}{Hongjun Choi}, \bibinfo{person}{Xiangyu Zhang}, {and} \bibinfo{person}{Sebastian~G. Elbaum}.} \bibinfo{year}{2021}\natexlab{}.
\newblock \showarticletitle{{PHYSFRAME:} type checking physical frames of reference for robotic systems}. In \bibinfo{booktitle}{\emph{{ESEC/FSE} '21: 29th {ACM} Joint European Software Engineering Conference and Symposium on the Foundations of Software Engineering, 2021}}. \bibinfo{publisher}{{ACM}}, \bibinfo{pages}{45--56}.
\newblock
\urldef\tempurl%
\url{https://doi.org/10.1145/3468264.3468608}
\showDOI{\tempurl}


\bibitem[Kim and Kim(2022)]%
        {DBLP:conf/sigsoft/KimK22-RoboFuzz}
\bibfield{author}{\bibinfo{person}{Seulbae Kim} {and} \bibinfo{person}{Taesoo Kim}.} \bibinfo{year}{2022}\natexlab{}.
\newblock \showarticletitle{RoboFuzz: fuzzing robotic systems over robot operating system {(ROS)} for finding correctness bugs}. In \bibinfo{booktitle}{\emph{Proceedings of the 30th {ACM} Joint European Software Engineering Conference and Symposium on the Foundations of Software Engineering, {ESEC/FSE} 2022}}. \bibinfo{publisher}{{ACM}}, \bibinfo{pages}{447--458}.
\newblock
\urldef\tempurl%
\url{https://doi.org/10.1145/3540250.3549164}
\showDOI{\tempurl}


\bibitem[Kim et~al\mbox{.}(2024)]%
        {DBLP:journals/isrob/KimKCPOP24-motionplanning}
\bibfield{author}{\bibinfo{person}{Yeseung Kim}, \bibinfo{person}{Dohyun Kim}, \bibinfo{person}{Jieun Choi}, \bibinfo{person}{Jisang Park}, \bibinfo{person}{Nayoung Oh}, {and} \bibinfo{person}{Daehyung Park}.} \bibinfo{year}{2024}\natexlab{}.
\newblock \showarticletitle{A survey on integration of large language models with intelligent robots}.
\newblock \bibinfo{journal}{\emph{Intell. Serv. Robotics}} \bibinfo{volume}{17}, \bibinfo{number}{5} (\bibinfo{year}{2024}), \bibinfo{pages}{1091--1107}.
\newblock
\urldef\tempurl%
\url{https://doi.org/10.1007/S11370-024-00550-5}
\showDOI{\tempurl}


\bibitem[Kolak et~al\mbox{.}(2020a)]%
        {DBLP:conf/icsm/KolakAGHT20-ecosystem}
\bibfield{author}{\bibinfo{person}{Sophia Kolak}, \bibinfo{person}{Afsoon Afzal}, \bibinfo{person}{Claire {Le Goues}}, \bibinfo{person}{Michael Hilton}, {and} \bibinfo{person}{Christopher~Steven Timperley}.} \bibinfo{year}{2020}\natexlab{a}.
\newblock \showarticletitle{It Takes a Village to Build a Robot: An Empirical Study of The {ROS} Ecosystem}. In \bibinfo{booktitle}{\emph{{IEEE} International Conference on Software Maintenance and Evolution, {ICSME} 2020}}. \bibinfo{publisher}{{IEEE}}, \bibinfo{pages}{430--440}.
\newblock
\urldef\tempurl%
\url{https://doi.org/10.1109/ICSME46990.2020.00048}
\showDOI{\tempurl}


\bibitem[Kolak et~al\mbox{.}(2020b)]%
        {DBLP:conf/icsm/KolakAGHT20-village}
\bibfield{author}{\bibinfo{person}{Sophia Kolak}, \bibinfo{person}{Afsoon Afzal}, \bibinfo{person}{Claire {Le Goues}}, \bibinfo{person}{Michael Hilton}, {and} \bibinfo{person}{Christopher~Steven Timperley}.} \bibinfo{year}{2020}\natexlab{b}.
\newblock \showarticletitle{It Takes a Village to Build a Robot: An Empirical Study of The {ROS} Ecosystem}. In \bibinfo{booktitle}{\emph{{IEEE} International Conference on Software Maintenance and Evolution, {ICSME} 2020, 2020}}. \bibinfo{publisher}{{IEEE}}, \bibinfo{pages}{430--440}.
\newblock
\urldef\tempurl%
\url{https://doi.org/10.1109/ICSME46990.2020.00048}
\showDOI{\tempurl}


\bibitem[Luckcuck et~al\mbox{.}(2022)]%
        {luckcuck2022compositional-modular-robotic}
\bibfield{author}{\bibinfo{person}{Matt Luckcuck}, \bibinfo{person}{Marie Farrell}, \bibinfo{person}{Angelo Ferrando}, \bibinfo{person}{Rafael~C Cardoso}, \bibinfo{person}{Louise~A Dennis}, {and} \bibinfo{person}{Michael Fisher}.} \bibinfo{year}{2022}\natexlab{}.
\newblock \showarticletitle{A compositional approach to verifying modular robotic systems}.
\newblock \bibinfo{journal}{\emph{arXiv preprint arXiv:2208.05507}} (\bibinfo{year}{2022}).
\newblock


\bibitem[Malavolta et~al\mbox{.}(2020)]%
        {DBLP:conf/icse/MalavoltaLSLG20-architect}
\bibfield{author}{\bibinfo{person}{Ivano Malavolta}, \bibinfo{person}{Grace~A. Lewis}, \bibinfo{person}{Bradley~R. Schmerl}, \bibinfo{person}{Patricia Lago}, {and} \bibinfo{person}{David Garlan}.} \bibinfo{year}{2020}\natexlab{}.
\newblock \showarticletitle{How do you architect your robots?: state of the practice and guidelines for ROS-based systems}. In \bibinfo{booktitle}{\emph{{ICSE-SEIP} 2020: 42nd International Conference on Software Engineering, Software Engineering in Practice}}. \bibinfo{publisher}{{ACM}}, \bibinfo{pages}{31--40}.
\newblock
\urldef\tempurl%
\url{https://doi.org/10.1145/3377813.3381358}
\showDOI{\tempurl}


\bibitem[Metta et~al\mbox{.}(2006)]%
        {metta2006yarp}
\bibfield{author}{\bibinfo{person}{Giorgio Metta}, \bibinfo{person}{Paul Fitzpatrick}, {and} \bibinfo{person}{Lorenzo Natale}.} \bibinfo{year}{2006}\natexlab{}.
\newblock \showarticletitle{YARP: yet another robot platform}.
\newblock \bibinfo{journal}{\emph{International Journal of Advanced Robotic Systems}} \bibinfo{volume}{3}, \bibinfo{number}{1} (\bibinfo{year}{2006}), \bibinfo{pages}{8}.
\newblock


\bibitem[Patil(2024)]%
        {Recovery-bahavior}
\bibfield{author}{\bibinfo{person}{Aditya Patil}.} \bibinfo{year}{2024}\natexlab{}.
\newblock \bibinfo{booktitle}{\emph{Recovery Behaviours used in Navigation Stack}}.
\newblock
\urldef\tempurl%
\url{https://medium.com/@patiladitya1309/recovery-behaviours-used-in-navigation-stack-ros-5df991f408c7}
\showURL{%
Retrieved March 7, 2025 from \tempurl}


\bibitem[Quigley et~al\mbox{.}(2009)]%
        {quigley2009ros}
\bibfield{author}{\bibinfo{person}{Morgan Quigley}, \bibinfo{person}{Ken Conley}, \bibinfo{person}{Brian Gerkey}, \bibinfo{person}{Josh Faust}, \bibinfo{person}{Tully Foote}, \bibinfo{person}{Jeremy Leibs}, \bibinfo{person}{Rob Wheeler}, \bibinfo{person}{Andrew~Y Ng}, {et~al\mbox{.}}} \bibinfo{year}{2009}\natexlab{}.
\newblock \showarticletitle{ROS: an open-source Robot Operating System}. In \bibinfo{booktitle}{\emph{ICRA workshop on open source software}}, Vol.~\bibinfo{volume}{3}. Kobe, Japan, \bibinfo{pages}{5}.
\newblock


\bibitem[Robotics(2019)]%
        {docs-executor}
\bibfield{author}{\bibinfo{person}{Open Robotics}.} \bibinfo{year}{2019}\natexlab{}.
\newblock \bibinfo{booktitle}{\emph{Executors}}.
\newblock
\urldef\tempurl%
\url{https://docs.ros.org/en/humble/Concepts/Intermediate/About-Executors.html#executors}
\showURL{%
Retrieved March 2, 2025 from \tempurl}


\bibitem[Robotics(2025)]%
        {ROS-Index}
\bibfield{author}{\bibinfo{person}{Open Robotics}.} \bibinfo{year}{2025}\natexlab{}.
\newblock \bibinfo{booktitle}{\emph{ROS Index}}.
\newblock
\urldef\tempurl%
\url{https://index.ros.org/repos/}
\showURL{%
Retrieved January 11, 2025 from \tempurl}


\bibitem[ros visualization(2016)]%
        {Rqt}
\bibfield{author}{\bibinfo{person}{ros visualization}.} \bibinfo{year}{2016}\natexlab{}.
\newblock \bibinfo{booktitle}{\emph{rqt}}.
\newblock
\urldef\tempurl%
\url{https://github.com/ros-visualization/rqt}
\showURL{%
Retrieved February 27, 2025 from \tempurl}


\bibitem[Scott and Foote(2024)]%
        {2023-ROS-Metrics-Report}
\bibfield{author}{\bibinfo{person}{Katherine Scott} {and} \bibinfo{person}{Tully Foote}.} \bibinfo{year}{2024}\natexlab{}.
\newblock \bibinfo{booktitle}{\emph{2023 ROS Metrics Report}}.
\newblock
\urldef\tempurl%
\url{https://discourse.ros.org/t/2023-ros-metrics-report/35837}
\showURL{%
Retrieved October 26, 2024 from \tempurl}


\bibitem[Seaman(1999)]%
        {DBLP:journals/tse/Seaman99-OpenCodingProcedure}
\bibfield{author}{\bibinfo{person}{Carolyn~B. Seaman}.} \bibinfo{year}{1999}\natexlab{}.
\newblock \showarticletitle{Qualitative Methods in Empirical Studies of Software Engineering}.
\newblock \bibinfo{journal}{\emph{{IEEE} Trans. Software Eng.}} \bibinfo{volume}{25}, \bibinfo{number}{4} (\bibinfo{year}{1999}), \bibinfo{pages}{557--572}.
\newblock
\urldef\tempurl%
\url{https://doi.org/10.1109/32.799955}
\showDOI{\tempurl}


\bibitem[Shen et~al\mbox{.}(2024)]%
        {DBLP:conf/issta/ShenLXSWGS024-FuzzingCallback}
\bibfield{author}{\bibinfo{person}{Yuheng Shen}, \bibinfo{person}{Jianzhong Liu}, \bibinfo{person}{Yiru Xu}, \bibinfo{person}{Hao Sun}, \bibinfo{person}{Mingzhe Wang}, \bibinfo{person}{Nan Guan}, \bibinfo{person}{Heyuan Shi}, {and} \bibinfo{person}{Yu Jiang}.} \bibinfo{year}{2024}\natexlab{}.
\newblock \showarticletitle{Enhancing {ROS} System Fuzzing through Callback Tracing}. In \bibinfo{booktitle}{\emph{Proceedings of the 33rd {ACM} {SIGSOFT} International Symposium on Software Testing and Analysis, {ISSTA} 2024}}. \bibinfo{publisher}{{ACM}}, \bibinfo{pages}{76--87}.
\newblock
\urldef\tempurl%
\url{https://doi.org/10.1145/3650212.3652111}
\showDOI{\tempurl}


\bibitem[Song et~al\mbox{.}(2023)]%
        {DBLP:conf/icsm/SongLDLCP23-FaultDiagnosis}
\bibfield{author}{\bibinfo{person}{Xuezhi Song}, \bibinfo{person}{Yi Li}, \bibinfo{person}{Zhen Dong}, \bibinfo{person}{Shuning Liu}, \bibinfo{person}{Junming Cao}, {and} \bibinfo{person}{Xin Peng}.} \bibinfo{year}{2023}\natexlab{}.
\newblock \showarticletitle{An Empirical Study on Fault Diagnosis in Robotic Systems}. In \bibinfo{booktitle}{\emph{{IEEE} International Conference on Software Maintenance and Evolution, {ICSME} 2023, 2023}}. \bibinfo{publisher}{{IEEE}}, \bibinfo{pages}{207--219}.
\newblock
\urldef\tempurl%
\url{https://doi.org/10.1109/ICSME58846.2023.00030}
\showDOI{\tempurl}


\bibitem[Tang et~al\mbox{.}(2023)]%
        {DBLP:conf/eurosys/TangBZKJGX23}
\bibfield{author}{\bibinfo{person}{Lilia Tang}, \bibinfo{person}{Chaitanya Bhandari}, \bibinfo{person}{Yongle Zhang}, \bibinfo{person}{Anna Karanika}, \bibinfo{person}{Shuyang Ji}, \bibinfo{person}{Indranil Gupta}, {and} \bibinfo{person}{Tianyin Xu}.} \bibinfo{year}{2023}\natexlab{}.
\newblock \showarticletitle{Fail through the Cracks: Cross-System Interaction Failures in Modern Cloud Systems}. In \bibinfo{booktitle}{\emph{Proceedings of the Eighteenth European Conference on Computer Systems, EuroSys 2023, 2023}}. \bibinfo{publisher}{{ACM}}, \bibinfo{pages}{433--451}.
\newblock
\urldef\tempurl%
\url{https://doi.org/10.1145/3552326.3587448}
\showDOI{\tempurl}


\bibitem[Timperley et~al\mbox{.}(2022)]%
        {DBLP:conf/icsa/TimperleyDSGG22-ROSDiscover}
\bibfield{author}{\bibinfo{person}{Christopher~Steven Timperley}, \bibinfo{person}{Tobias D{\"{u}}rschmid}, \bibinfo{person}{Bradley~R. Schmerl}, \bibinfo{person}{David Garlan}, {and} \bibinfo{person}{Claire {Le Goues}}.} \bibinfo{year}{2022}\natexlab{}.
\newblock \showarticletitle{ROSDiscover: Statically Detecting Run-Time Architecture Misconfigurations in Robotics Systems}. In \bibinfo{booktitle}{\emph{19th {IEEE} International Conference on Software Architecture, {ICSA} 2022}}. \bibinfo{publisher}{{IEEE}}, \bibinfo{pages}{112--123}.
\newblock
\urldef\tempurl%
\url{https://doi.org/10.1109/ICSA53651.2022.00019}
\showDOI{\tempurl}


\bibitem[Timperley et~al\mbox{.}(2024)]%
        {DBLP:journals/ese/TimperleyHSDW24-ROBUST221bugs}
\bibfield{author}{\bibinfo{person}{Christopher~Steven Timperley}, \bibinfo{person}{Gijs van~der Hoorn}, \bibinfo{person}{Andr{\'{e}} Santos}, \bibinfo{person}{Harshavardhan Deshpande}, {and} \bibinfo{person}{Andrzej Wasowski}.} \bibinfo{year}{2024}\natexlab{}.
\newblock \showarticletitle{{ROBUST:} 221 bugs in the Robot Operating System}.
\newblock \bibinfo{journal}{\emph{Empir. Softw. Eng.}} \bibinfo{volume}{29}, \bibinfo{number}{3} (\bibinfo{year}{2024}), \bibinfo{pages}{57}.
\newblock
\urldef\tempurl%
\url{https://doi.org/10.1007/S10664-024-10440-0}
\showDOI{\tempurl}


\bibitem[Xie et~al\mbox{.}(2022)]%
        {DBLP:conf/icra/XieBZ022-ROZZ-Fuzz}
\bibfield{author}{\bibinfo{person}{Kai{-}Tao Xie}, \bibinfo{person}{Jia{-}Ju Bai}, \bibinfo{person}{Yong{-}Hao Zou}, {and} \bibinfo{person}{Yu{-}Ping Wang}.} \bibinfo{year}{2022}\natexlab{}.
\newblock \showarticletitle{{ROZZ:} Property-based Fuzzing for Robotic Programs in {ROS}}. In \bibinfo{booktitle}{\emph{2022 International Conference on Robotics and Automation, {ICRA} 2022, P 2022}}. \bibinfo{publisher}{{IEEE}}, \bibinfo{pages}{6786--6792}.
\newblock
\urldef\tempurl%
\url{https://doi.org/10.1109/ICRA46639.2022.9811701}
\showDOI{\tempurl}


\end{thebibliography}

\end{document}